\documentclass[twoside,twocolumn]{article}

\usepackage[sc]{mathpazo} 

\usepackage[T1]{fontenc} 
\usepackage{lmodern} 
\usepackage{microtype} 
\usepackage[version=4]{mhchem}
\usepackage[english]{babel} 

\usepackage{graphicx}
\usepackage{color}
\usepackage[top=2cm,bottom=2cm,left=1.5cm,right=1.5cm,columnsep=20pt]{geometry} 
    {\large\scshape Computational Details%
    \par\medskip\normalfont\normalsize}%
    {}%
    {\large\scshape Acknowledgement%
    \par\medskip\normalfont\normalsize}%
    {}%
    {\large\scshape Supporting Information%
    \par\medskip\normalfont\normalsize}%
    {}%
\usepackage{amsmath}
\usepackage{amssymb}
\usepackage{multirow} 
\usepackage[normal, small,labelfont=bf,up,textfont=it,up]{caption} 
\usepackage{booktabs} 

\usepackage{lettrine} 

\usepackage{enumitem} 
\setlist[itemize]{noitemsep} 

\usepackage{abstract} 

\usepackage{titlesec} 
\renewcommand\thesection{\Roman{section}} 
\renewcommand\thesubsection{\roman{subsection}} 
\titleformat{\section}[block]{\large\scshape\centering}{\thesection.}{1em}{} 
\titleformat{\subsection}[block]{\large}{\thesubsection.}{1em}{} 

\usepackage{fancyhdr} 
\usepackage{titling} 

\usepackage{hyperref} 

\pretitle{\begin{center}\Large\bfseries} 
\posttitle{\end{center}} 

\def\bea{\begin{eqnarray}}
\def\eea{\end{eqnarray}}
\def\ben{\begin{equation}}
\def\een{\end{equation}}
\def\benu{\begin{enumerate}}
\def\enu{\end{enumerate}}

\def\bei{\begin{itemize}}
\def\eei{\end{itemize}}
\def\beit{\begin{itemize}}
\def\eit{\end{itemize}}
\def\benu{\begin{enumerate}}
\def\enu{\end{enumerate}}

\def\sss{\scriptscriptstyle\rm}

\def\1var{(\bx_1...\bx\N)}

\def\half{\frac{1}{2}}

\def\bx{{x}}

\def\N{_{\sss N}}
\def\H{_{\sss H}}

\def\sph_int{ {\int d^3 r}}

\usepackage{caption}
\usepackage{float}
\usepackage{graphicx,subcaption}
\usepackage{chemformula}
\usepackage[version=4]{mhchem}

\usepackage{xr}
\usepackage{float}
\usepackage{array,booktabs}
\usepackage{hyperref}
\usepackage{xcolor}
\usepackage[normalem]{ulem}
\useunder{\uline}{\ulined}{}%
\DeclareUrlCommand{\burl}{}

\title{Quantifying and understanding errors in molecular geometries}

\author{%
\textsc{Stefan Vuckovic and Kieron Burke}\thanks{kieron@uci.edu} \\  \\ 
\normalsize Departments of Chemistry and of Physics, University of California, Irvine, CA 92697, USA \\  
}
\date{\today} 

\renewcommand{\maketitlehookd}{
\begin{abstract}
\noindent 
Electronic structure calculations are ubiquitous in most branches of chemistry, but
all have errors in both energies and equilibrium geometries.  
Quantifying errors in possibly dozens of bond angles and bond lengths is a Herculean task.
A single natural measure of geometric error is introduced, the geometry energy offset (GEO).
GEO links many disparate aspects of geometry errors:
a new ranking of different methods, quantitative insight into 
errors in specific geometric parameters, and insight into trends with different 
methods.
GEO can also reduce the cost of high-level geometry optimizations and shows when
geometric errors 
distort the overall error of a method.
Results, including some surprises, are given for both covalent and weak interactions.
\end{abstract}
}

\def\beq{\begin{equation}}
\def\eeq{\end{equation}}

\def\half{\frac{1}{2}}

\def\tE{\tilde E}

\def\tE{\tilde E}
\def\G{\mathbf{G}}

\def\tG{\tilde {\mathbf{G}}}

\def\H{\mathbf{H}}

\def\geo{E_{geo}}
\def\geoh{E_{geo}^{\rm harm}}
\def\geos{E_{geo}^{\rm simple}}
\def\geop{E_{geo}^{\prime}}

\def\K{D}
\def\geod{E_{geo}^D}

\begin{document}

\maketitle
\chapter{}
\setcounter{secnumdepth}{2}
\renewcommand\thesubsection{\Alph{subsection}}

\sf
Whenever one runs an electronic structure calculation within the Born-Oppenheimer approximation, whether it is
a density functional calculation, {\em ab initio} or semiempirical, of a molecule or a material, one must always answer the
question:  Which geometry should I use?   Whatever the limitations of your method are, they will
show up in giving an approximate energy at any given geometry which will minimize at some approximate geometry.
Sometimes the differences between the true geometry and the approximation are so slight that it does not matter.
Whenever it does matter, common sense often dictates a choice:  When comparing different methods,
the requirement of apples-to-apples comparison means comparing several methods
with a fixed geometry.~\cite{GoeHanBauEhrNajGri-PCCP-17,YuZhaBerHeTru-PCCP-15,MarHea-MP-17}
Other times, when the cost of a single calculation is severe, geometry optimization is prohibitively
expensive, and one resorts to using geometries from a cheaper method.

This problem is compounded when comparing geometric parameters computed with different methods.
As a molecule grows in size, there are $3N-6$ distinct degrees of freedom for the equilibrium
structure, with errors in bond lengths, angles, etc.  Some are more accurate in one method,
some are better in another (see, e.g., Ref.~\cite{SX16}).  Should one average over all such parameters?  But what if one
method is better for bond lengths, and another for angles?  And how do such errors in geometry
correlate with other energetic errors?

\begin{figure*}[htb!]
\centering
\includegraphics[width=2\columnwidth]{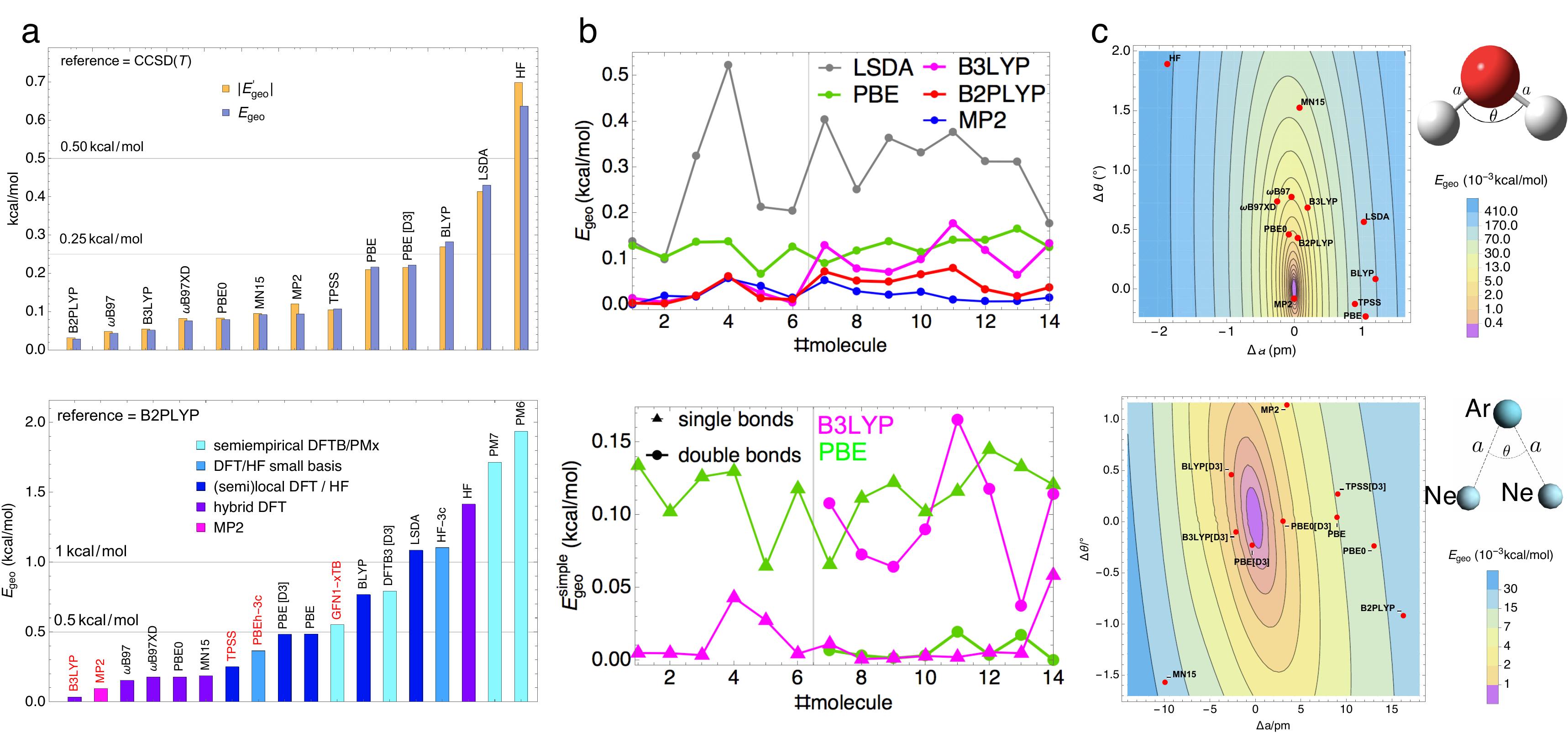}
\caption{Why GEO is useful.  
Left:  
Exact and approximate
GEO rankings for quantum chemical methods on small molecules (top), and
GEO of many different methods over medium-sized
organic molecules (bottom) with B2PLYP as reference.
For the lists of molecules and further details, see Figs.~S2-S6, and Tables~S1-S3.
Center:  
$\geo$ for a few methods and a few molecules (top), single and
double-bond contributions to $\geos$ for
two popular methods, showing huge difference in their accuracies (bottom). For the list of molecules, see Fig.~\ref{fig:K}.
Right: GEO contours as a function of errors in the bond angle and length for the water molecule (top panel) and weakly bonded Ne$_2$Ar (lower panel), and positions of different approximations.}
\label{fig:f}
\end{figure*}

We define the geometry energy offset of a given method as
\ben \label{eq:err1}
\geo = E \left (\tG\right )  - E\left (\G_0\right )
\een
where $E(\G)$ is the ground-state energy at geometry $\G$, $\G_0$ the exact geometry and $\tG$ an approximate geometry.
This simple definition leads to all the analysis and results contained in the paper. 
Figure \ref{fig:f} summarizes some of our most important results with GEO, with more details within this
paper and supplementary information.   On the left,
we plot GEO energies averaged over a data set of small organic molecules (top).
Accurate calculation of GEO using CCSD(T) is expensive (see methods), but using the approximations themselves is
less expensive and yields nearly identical results ($\geop$).
We can see that different approximations perform characteristically well or poorly.  Thus every method
can be ranked by its GEO value, and some perform much better for geometries than they do for, e.g., atomization
energies.  This is crucial information for understanding the accuracy of different methods for geometries.
Directly below, we evaluate a much greater variety of methods for a data set
of medium-sized organic molecules.  Here, we use
B2PLYP as the reference, since it is the winner
in the top panel and CCSD(T) is already too expensive.  
This shows some surprises: lower level (and less costly)
methods can outperform higher level methods because they have been trained empirically.   For example
the semiempirical GFN1-xTB method of Grimme and co-workers competes with
DFT with the PBE functional\cite{PerBurErn-PRL-96}, and outperforms
DFT with BLYP\cite{Bec-PRA-88,LeeYanPar-PRB-88}.

In the center, we show the very disparate
behavior of two popular representative density functionals for single bonds and for double bonds.  PBE, as a generalized gradient approximation (GGA) is far more accurate for double bonds than for singles.  But a
(global) hybrid, B3LYP, totally reverses this trend:  
more accurate for single bonds,
but surprisingly far less
accurate for double bonds.  Explanations of the accuracy of hybrids\cite{B93,BurErnPer-CPL-97} typically center on atomization
energies, not bond lengths, and do not explain these trends.
The top right panel shows a trade-off between angle- and bond-length errors.
There are clear behaviors of different levels of density functionals.  The local density approximation has noticeable
errors in both the bond length and angle (but far smaller than those of HF).  
One can clearly see how GGA's like PBE and BLYP and the meta-GGA TPSS~\cite{TaoPerStaScu-PRL-03} greatly reduce the angular error, but
have almost no effect on the bond length.  Finally, by mixing some fraction (about 1/4) of exact exchange,
PBE0 and B3LYP lie along a line joining their parent GGA to HF, and the mixing fraction almost perfectly cancels
the bond-length error, while increasing the angle-error.  Better functionals have about the same accuracy,
while MP2 has almost perfect geometry.  

All these results and trends are for strong covalent bonds.  But GEO is even more important for weak
interactions, where GEO energies can be comparable to the binding energy itself.
To illustrate this, in the right, we contrast contours of GEO for two A$_2$B molecules, one covalently bonded and the other
a non-covalent interaction:  Water and the van der Waals trimer, Ne$_2$Ar,
with the different methods from the left figures plotted as points in the plane.  
The non-covalent case is strikingly different.  First, its binding energy is only 0.37 kcal/mol, so GEO errors
are now more than relevant on this scale.  This is accompanied by huge errors in bond length, related to
the softness of the potential.  Finally, the performance of different electronic structure methods is 
very different from the covalent case.  For covalent bonds, MP2 is exceptionally good; for NCI's, approximate
density functionals are much better.   These effects seem to have largely been ignored when ranking
functionals for such complexes, which is usually done at a fixed geometry.~\cite{GoeHanBauEhrNajGri-PCCP-17,YuZhaBerHeTru-PCCP-15,MarHea-MP-17}  We expect improved
performance for weak interaction methods once GEO errors are accounted for.

The rest of this paper explains how GEO works and shows how useful it can be.
We focus on just three immediate applications:
(i) obtaining insight into geometric errors in molecular benchmark energy sets;
(ii) establishing an energetic scale comparing the quality of geometries from
different approximate quantum-mechanical (QM) solvers;
(iii) how this scale can be used for choosing a geometry optimization solver that
has a good accuracy to cost ratio.
We apply our logic first to covalent bonds, where GEO is typically negligible relative to atomization
energies, and then to non-covalent bonds, where GEO is often comparable to binding energies, and so is even
more important.

\noindent
{\bf Results and Discussion}

{\bf Performance of approximations.}
One of the most valuable uses of GEO is to rank different electronic structure methods for
their geometric accuracy, as illustrated in the upper left panel of Fig 1.  We stress that
this ranking is quite different from traditional rankings by purely energetic performance,
such as for atomization energies (AE). In Fig.~S2 (top panel), we give the errors for AE
on our set of small organic molecules, and in Fig.~S2 (lower panel), we show the correlation plot between GEO and MAE for AE.
Roughly, GEO is typically about 1/80-th of the mean absolute error in
AE. By such a measure, LSDA and HF are surprisingly good for geometries, because
their MAE for atomization is so poor.  Non-empirical functionals (PBE, TPSS) are close to this
line, so improvements in AE's are reflected in improvements in GEO.  Empirical functionals typically
perform very well, especially the global hybrid B3LYP, and BLYP, as a GGA, yields
surprisingly poor geometries.  Also, the
addition of D3 corrections to any
functional makes little difference to its geometric performance for covalent bonds.

If we want to calculate GEO errors for larger main-group molecules, 
for which CCSD(T) is too expensive, we can use B2PLYP as a reference in place of CCSD(T). 
Multireference systems, such as transition metal dimers, are more delicate and would need a better reference than CCSD(T)~\cite{XuZhaTanTru-JCTC-15,HaiTubLevWhaHea-JCTC-19} to calculate GEO.

The bigger picture is shown in the lower-left panel of Fig. 1, which includes many different
kinds of methods.  Here, we had to use B2PLYP~\cite{f_B2PLYP} as a reference (see above).
Besides the QM solvers considered in Figure~\ref{fig:f}a(top),
we also include semiempirical QM solvers, such as DFTB\cite{DFTB3,GFN1-xTB} and PM$x$\cite{PM6,PM7},
and the highly practical HF-3c\cite{HF-3c} and
PBEh-3c\cite{PBEh-3c}, both of which use a small basis set and contain empirical parameters.
The results are summarized in the lower panel of Figure~\ref{fig:f}a,
where
different  error bar colours indicate methods at different levels of theory,
with the best method for each level of theory shown in red.
Trends
are similar to the panel above, but overall GEO's are larger as the molecules are bigger.
This plot is not accurate below about 0.1 kcal/mol, because of the
B2PLYP reference.  Thus B3LYP does not really rank as  No. 1, its errors are simply correlated
with the reference.  As we discuss later, since
GFN1-xTB\cite{GFN1-xTB} has the best  performance among all semiempirical
methods shown, it can serve as an excellent starting point in optimization schemes. 

{\bf Geometry optimization.}
Based on GEO, one can establish the following {\em sequence} composed of
 the best method for each level of complexity: GFN1-xTB $\to$ TPSS 
(or PBEh-3c) $\to$ B3LYP $\to$ B2PLYP. This sequence can be used in 
automated explorations of chemical space and 
molecular screenings assisted by QM solvers,\cite{Gri-JCTC-19,nanoreactor,CREST} which are powerful tools 
for the discovery of new molecules with desired properties. 
In these procedures, based on energetic criteria  (e.g., their binding energies with
a specific enzyme) molecules are filtered out, and 
QM geometry optimizations of a large number of molecules make the procedure computationally demanding.
As the number of molecules in the screening decreases,
more expensive and more accurate methods are typically used. Thus, based on our sequence determined by the GEO criterion, in the first step of the screening one can 
employ GFN1-xTB for optimizing geometries of all initial molecular candidates. 
After the first cycle of filtering out molecules, TPSS\cite{TaoPerStaScu-PRL-03} can be
employed as an optimizer, and so on. In the last round, B2PLYP geometries can
be confidently used, given they are energetically very close to
the CCSD(T) ones ($\sim$0.03 kcal/mol for the testset considered in Figure~\ref{fig:f}a). 
Even if one only wants the CCSD(T) geometries for small molecules,  one can use the
same {\em sequence} to pre-optimize the molecular geometry, before the
CCSD(T) optimizer  is turned on, and thereby save computational time.

\begin{figure*}[htb!]
\includegraphics[width=2\columnwidth]{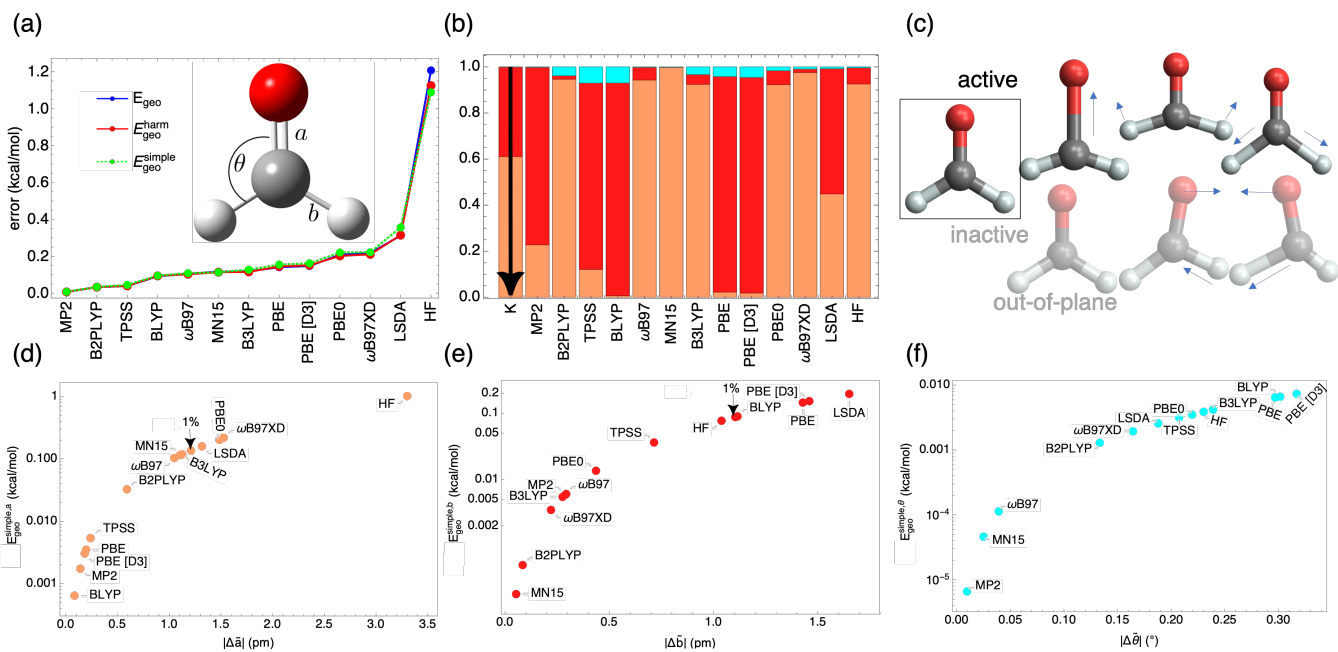}
\caption{GEO analysis for formaldehyde.
Top: (a) GEO rankings of approximations. The plots also show that $\geo$ is accurately approximated by $\geoh$ (Eq.~\ref{eq:ga}) and $\geos$ (Eq.~\ref{eq:gas}).  
(b) $\geos$ weights, $E_{geo}^{\rm simple,i}/ \geos$ (see lower panels for colour legends). When all bonds are stretched by the same $\gamma$ factor, GEO becomes $\geo^\gamma$ (see the text), and the  underlying weights are marked by the arrow. 
(c) GEO-active and GEO-inactive modes (those that have no contribution to the r.h.s. of Eq~\ref{eq:gaNORM}). For the $\geoh$ weights in normal modes, analogous to panel~(b), see Fig.~S28. 
(d)-(f) Plots showing errors in individual geometric parameters (x-axis), and how these errors translate to $\geos$ terms by virtue of Eq.~\ref{eq:gas} (y-axis). The points marked by the arrows, show GEO when the bond lengths are stretched by 1$\%$.}
\label{fig:fo}
\end{figure*}

{\bf Simplifications and analysis tools.}
As mentioned above, a much less costly calculation is
\ben \label{eq:err2}
\geop = \tilde E \left (\tG\right )  - \tilde E \left(    \G_0      \right ),
\een
where $\tilde E$ is the approximate energy (see Supplementary Section 1 for more mathematical details).  This is typically an excellent approximation to GEO,
because even inaccurate methods such as HF yield reasonable vibrational frequencies.\cite{Pul-JCP-83}
From Figure~\ref{fig:f}a, the mean $\geo$ error is
in close agreement with its $\geop$ counterpart for all approximations (see also Tables~S1 and~S2 for $\geo$  and $\geop$ values for individual molecules). 

The next simplification is to approximate GEO
by expanding $E_0$ around its minimum to second order:
\ben \label{eq:ga}
\geoh= \half  \Delta \tilde{\G}^\intercal \H_0  \Delta \tilde{\G},
\een
where $\H_0$ is the Hessian at the minimum, composed of force constants and 
$\Delta \tilde{\G}= \tilde{\G}-\G_0$ is the error in specific geometric parameters (degrees of freedom)
that determine the relative positions of nuclei.  These could be simple Cartesians
or any other choice of coordinates.  Again, this is extremely accurate when approximated
with most electronic structure methods (see Figures~S17-~S22). Thus, we can use Eq.~\ref{eq:ga} for
further analysis and decomposition of GEO. 
One can easily diagonalize $\H_0$ and obtain GEO modes, $p$. In these coordinates,  Eq.~\ref{eq:ga} becomes:
\ben  \label{eq:gaNORM}
\geoh= \half \sum_{i}^{N_g} f_i^p \left(\Delta p_i \right)^2,
\een
where $f_i^p$ are the underlying force constants ($\H_0$ eigenvalues) and $\Delta p_i = \tilde p_i - p_i$
represent $\Delta \tilde{\G}$ written in terms of the errors in the GEO normal modes ($\H_0$ eigenvectors) and $N_g \leq 3N-6$ is the number of GEO {\em active} modes.
A highly appealing feature of Eq~\ref{eq:gaNORM} is that each term contributes positively to $\geoh$, which, in turn, allows us to obtain weights of each modes' contribution to the total GEO. 
In Figures~S27-~S29, for a set of small molecules we analyse how each of
the $N_g$ GEO-active modes contributes to the total  $\geoh$ for
different approximations and we also show the {\em GEO-inactive} modes, those that have no contribution to $\geoh$.
For example, by symmetry, no (sensible) electronic structure approximation gives unequal \ce{OH} bond lengths in the water molecule,
so the asymmetric stretch of the \ce{OH} bond is GEO-inactive (see Fig.~S5).
The higher the symmetry of a molecule, the fewer modes are GEO active.  For ethene,
all modes that distort its $D_{2h}$ symmetry (asymmetric and out-of-plane vibrations) are GEO-inactive,
so only 3 of its 12 modes are GEO-active, as shown in Fig.~S29.

Besides the GEO modes, a more chemically intuitive analysis in terms of bond lengths, angles, and torsion angles
of $\geo$ can be obtained by considering Eq.~\ref{eq:ga} in internal coordinates. Considering only
the underlying diagonal elements of $\H_0^q$ (the Hessian in internal coordinates), we find
the following {\em simple} approximation to $\geoh$:
\ben  \label{eq:gas}
\geoh \approx \geos= \half \sum_{i}^{3N-6} f_{i,i}^q \left(\Delta q_i \right)^2. 
\een
where $\Delta q_i = \tilde q_i - q$ are the errors in internal coordinates. 
While the r.h.s of Eq.~\ref{eq:gaNORM} is exactly equal to $\geoh$, this is not so for $\geos$, 
since the off-diagonal $\H_0^q$ are typically small but non-zero. 
For the organic molecules we consider here, $\geos$ is typically in good agreement with
both $\geoh$ and the "exact" $\geo$ (see Figures~S17-~S22).
This, in turn, allows us to {\em safely} use Eq.~\ref{eq:gas} to decompose $\geoh$ into
its positive contributions arising from errors in specific geometric parameters.  

In Figure~\ref{fig:fo}, we illustrate how a GEO analysis works for a simple case, formaldehyde. In panel~(a),
we give GEO rankings of different approximations for this molecule, which somewhat align 
with the database averages of Fig 1(a).
As with all covalent cases we studied, $\geoh$ and $\geos$ are in excellent agreement with GEO,
which allows us to use Eq~\ref{eq:gas} to decompose contributions from different structural parameters. 
The fractional contributions of each coordinate are shown in panel~(b). 
Angle errors give only a minor contribution to GEO for all methods. 
For the hybrids,  GEO error
comes nearly entirely from the error in the double bond, while
in the case of semilocal functionals, nearly
the entire GEO error comes from the error in the single bond lengths,
consistent with the trends in Fig 1(b).
The rankings for the single and double bond lengths and the bond angle are shown in the lower panels,
and how they correlate with $\geos$.  In each case, the actual curve is parabolic (the GEO axis is logarithmic).
The rankings differ substantially from those for the total GEO.
The leftmost is the double bond, and here the semilocal functionals (no mixing of exact exchange)
do best, outperforming even B2PLYP!   The hybrids do no better than simple LSDA.
But, the roles are reversed in the middle panel, showing that hybrids greatly
improve single-bond length error.  Finally, the angle-error is shown, with a
variety of results, but no clear trends.  The right panel of Fig.~S12 shows  that MP2 and B2PLYP yield the best angles on average for the set of molecules considered in Fig.~1b. All three curves are monotonic,
so rankings by a $\geos$ contribution correspond directly to rankings by the error in the underlying geometric parameter.

\begin{figure}[htb!]
\centering 
\includegraphics[width=1\columnwidth]{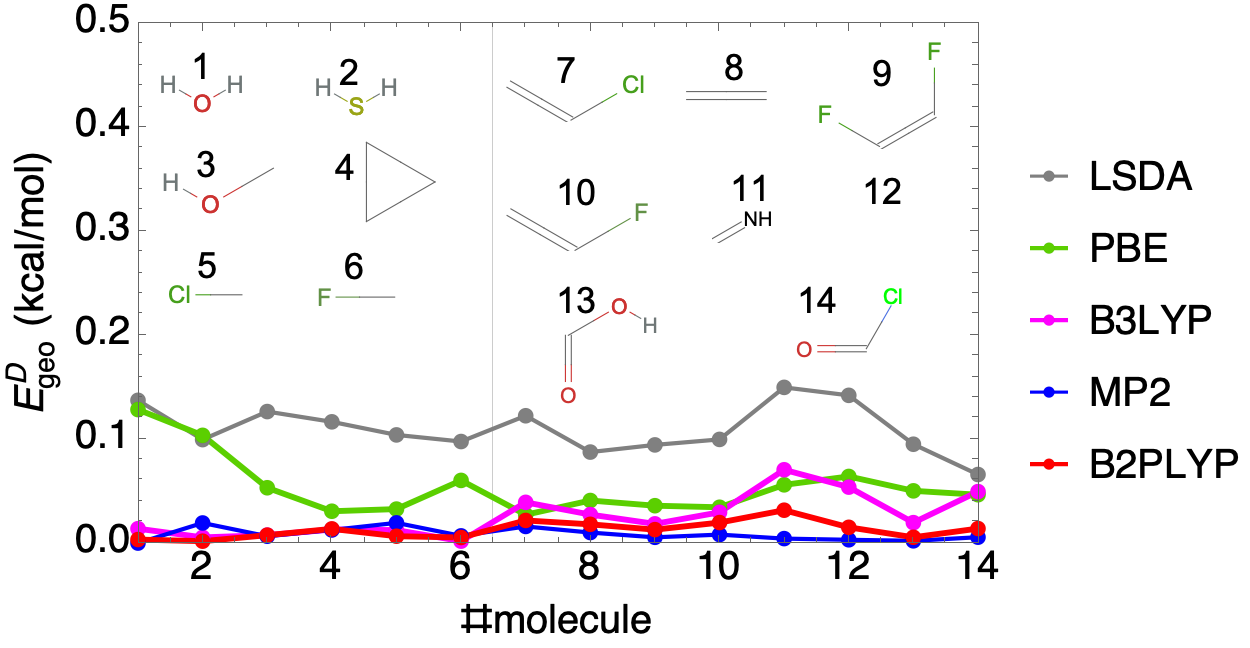}
\caption{The same plot as in the top panel of Fig.~\ref{fig:f}b but with $
\geod$ (see the absolute GEO scale section) on the $y$-axis. For more plots comparing $\geo$ and $\geod$, see~Fig.~S7.}
\label{fig:K}
\end{figure}

{\bf Absolute GEO scale.}
A problem that bedevils benchmarking of atomization energies is whether to
consider total energy errors or errors per bond.  Here we show that there
exists a universal GEO scale, independent of any method, that 
overcomes this problem for geometric errors.
Consider a small expansion of all coordinates, $\Delta \G= 
\gamma \G_0$, producing
$\geo^\gamma =\gamma^2  \K/2$ where $\K= \G_0^\intercal  \H_0  \G_0$.
Thus, $\geo^\gamma$ is the GEO value for a very specific geometric error, that
of expansion (or compression) of the exact geometry.  
For our small molecules, if $\gamma=1 \%$, $\geo^\gamma$ is a fraction of
a kcal/mol.  Thus, any calculation of GEO by any method for any molecule can be
compared to this intrinsic property of the molecule.  Moreover, $\geo^\gamma$
scales with the size of the molecule (compare, e.g., $\K$ values 
for small molecules shown in Table~S7 with those for medium-sized molecules shown in Table~S8), so that GEO's measured relative to it
do not grow with molecular size. We can even decompose $\geo^\gamma$ in terms
of Hessian eigenvectors or simple internal coordinates, giving an internally
defined distribution of contributions.  
This only includes bond lengths, as
no angle changes when molecules are uniformly expanded.  

We define: $\geod =\geo \left( \K_{{\rm H}_2 {\rm  O}}  / \K \right) $, and in Figure~\ref{fig:K}, we repeat the plot in the top panel of
Fig.~\ref{fig:f}b., but with $\geod$, which varies much less with
molecular size.  In Table~S6, we show that approximate calculations of
$\K$ typically yield highly accurate estimates (even HF is not too bad).
In Figure~S30, we show the decomposition of $\K$ for each bond in each
molecule, with double bonds being about 0.25 and most singles being about
0.1 times 10$^4$ kcal/mol if to an H atom, and about 0.16
if between heavier atoms. For rare gas dimers (bonded by weak interactions), $\K$ values are  several orders of magnitude smaller (see Table~S9).

Returning to Fig 2, the leftmost column of (b) is the $\K$-decomposition
of the single versus double bond, showing that under expansion, 60\% of
the energy cost is to stretch the double bond, 40\% to stretch the single.
Then BLYP clearly makes an unusually small error in the double, and a relatively
large error in the single.

\begin{table*}[!h]
\begin{tabular}{cccccccccc}
 \text{   } & \text{MP2/CBS} & \text{B2PLYP} & \text{B3LYP[D3]} & \text{PBE0[D3]} & \text{PBE[D3]} & \text{MN15} & \text{wB97XD} &
   \text{PBE} & \text{CCSD(T)} \\
 \text{MP2/CBS} & \textcolor{blue}{0.51} & 0.26 & 0.45 & 0.43 & 0.38 & 0.47 & 0.46 & 0.28 & \textcolor{red}{0.47} \\
 \text{B2PLYP} & 1.63 & \textcolor{blue}{1.34} & 1.45 & 1.43 & 1.38 & 1.52 & 1.51 & 1.40 &  \textcolor{red}{1.46} \\
 \text{B3LYP[D3]} & 0.44 & 0.34 & \textcolor{blue}{0.43} & 0.42 & 0.41 & 0.41 & 0.42 & 0.53 &  \textcolor{red}{0.43} \\
 \text{PBE0[D3]} & 0.51 & 0.43 & 0.48 & \textcolor{blue}{0.49} & 0.48 & 0.45 & 0.47 & 0.54 &  \textcolor{red}{0.48} \\
 \text{PBE[D3]} & 0.48 & 0.4 & 0.44 & 0.45 & \textcolor{blue}{0.46} & 0.40 & 0.41 & 0.44 &  \textcolor{red}{0.44} \\
 \text{MN15} & 0.56 & 0.34 & 0.59 & 0.58 & 0.53 & \textcolor{blue}{0.6} & 0.60 & 0.34 &  \textcolor{red}{0.59} \\
 \text{wB97XD} & 0.42 & 0.25 & 0.47 & 0.45 & 0.41 & 0.47 & \textcolor{blue}{0.48} & 0.38 &  \textcolor{red}{0.46} \\
 \text{PBE} & 2.29 & 1.77 & 2.04 & 1.98 & 1.90 & 2.14 & 2.12 & \textcolor{blue}{1.71} & \textcolor{red}{2.06} \\
\end{tabular}
\caption{MAE of different methods (rows) at different minima of S66x8 binding curves (columns) for the S66 dataset. The error is defined as: $E_X [R_Y] - E_0$, where $X$ is a method in rows and Y is the method in columns, and $E_0$ is the binding energy at the CCSD(T) minimum. The S66x8 binding curves from CCSD(T)/CBS (used as a reference here) and MP2/CBS have been taken from the original S66x8 dataset.\cite{RezRilHob-JCTC-11,rezac}  All other binding curves have been obtained from counterpoise corrected calculations within the aug-cc-pVTZ basis set. The minimum of each binding curve has been found numerically after the interpolation of 8 datapoints for each of the S66 complex (see details above). }
\label{tab_s66}
\end{table*}

\begin{figure}
\includegraphics[width=1\columnwidth]{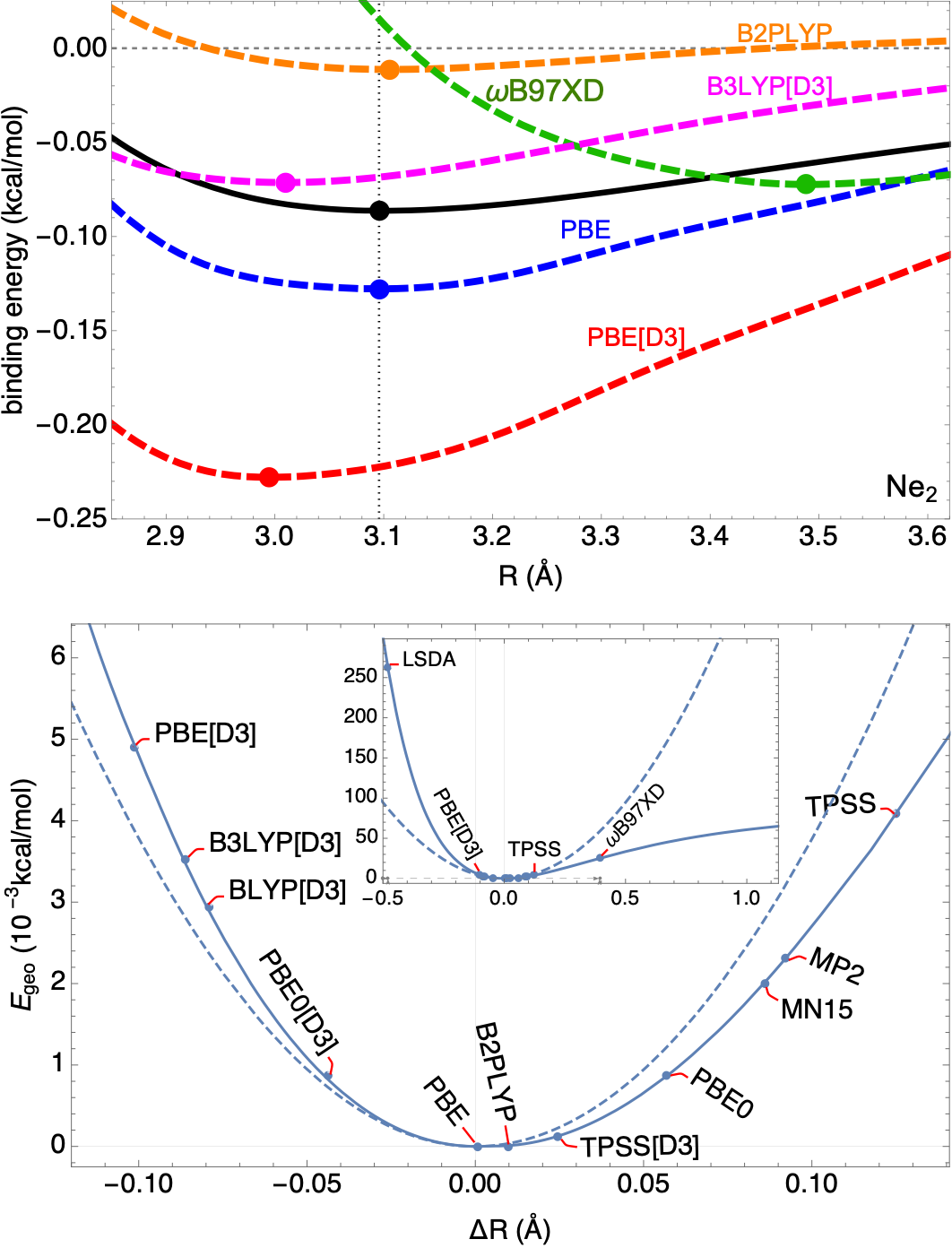}
\caption{GEO analysis for the \ce{Ne2} binding energies.
Top panel: Binding curves of \ce{Ne2}  with various methods. 
Lower panel: GEO for different approximations. For more details, see Figs.~S35-~S38.}
\label{fig:ne}
\end{figure}

\begin{figure}
\includegraphics[width=1\columnwidth]{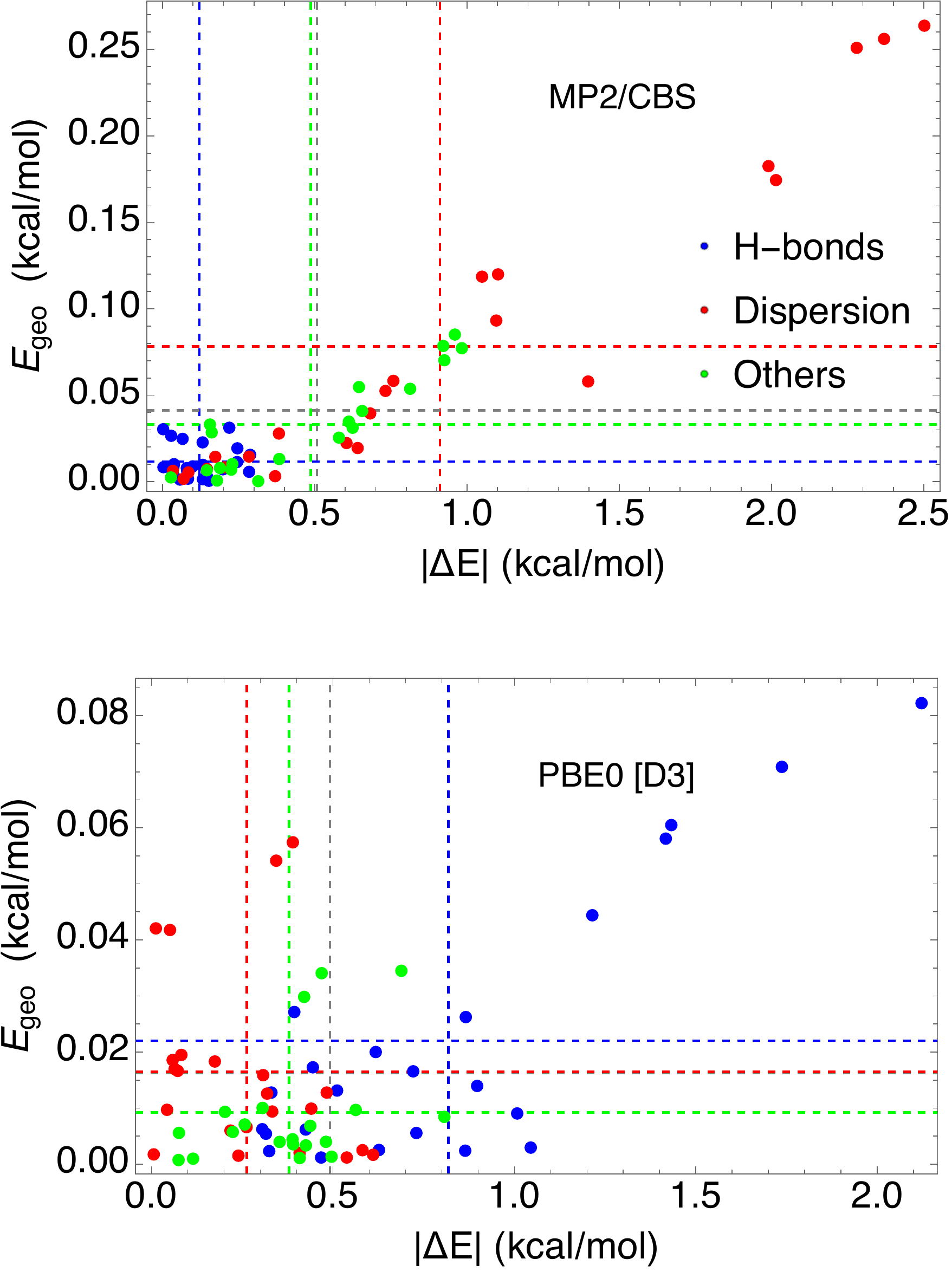}
\caption{The  $\Delta E=E_X [E_X] - E_0$ vs. GEO errors 
for the binding energies of the S66 complexes of
MP2/CBS (top panel), PBE0 [D3] (lower panel). The complexes are classified into "H-bonds", "Dispersion" and "Others" as in the original S66 publication.\cite{RezRilHob-JCTC-11}. Coloured dashed lines represent MAEs for different S66 categories and gray dash line represents the overall S66 MAEs.}
\label{fig:s66}
\end{figure}

{\bf Non-covalent interactions.}
The rest of this paper is devoted to weak interactions.  We re-examine all aspects of GEO
for these cases, as GEO energies can be a more significant fraction of the binding energies here.  Force constants for weak bonds are so much weaker that even very small GEO values can lead to large errors in bond lengths.
For weak interactions, we include D3 corrections~\cite{GriAntEhrKri-JCP-10} to the approximations, which typically greatly improve
energetic accuracy.  We also only allow the weak bond length to vary in complexes, i.e., one degree of freedom.

The upper panel shows a prototypical van der Waals system, Ne$_2$, with the exact curve and 
various approximations.  In most cases, the geometries
are quite accurate (minima are marked by beads) so that GEO energies are very small.
Surprisingly, $\omega$B97XD~\cite{fn_wB97X} has a large geometric error, but nonetheless yields
a highly accurate energy minimum, a cancellation of geometric and non-geometric errors (apparently
missed by its creators~\cite{fn_wB97X}).  Since functionals are applied to cases where neither accurate
geometries nor energies are known, the primary concern is to predict an accurate energy at the
approximate minimum.  By this criterion, $\omega$B97XD is the most accurate approximation shown!
Note that the D3 correction worsens PBE here.
Furthermore, B2PLYP gives an excellent geometry.

The approximate $\geop$ works well when the geometry is good, and
the harmonic approximation is largely still valid (dashed line in the lower panel of Fig.~\ref{fig:ne}), but is less accurate
than for typical covalent cases.  Decomposition into Hessian eigenvectors is irrelevant here
as we only adjust one geometry parameter, the length of the weak bond.  The analysis is thus
the same as for a single diatomic, where the cartesian coordinate difference is the internal coordinate.

Typically, databases are established using a fixed reference geometry, which may or may not be
very close to the true minimum (as measured by GEO).  Optimization of parameters in an approximation
will then miss the trade-off between geometry and energy that can occur in applications beyond
the training database, where presumably a geometry-optimized calculation should be designed to yield
the best energy.

{\bf Implications for benchmarking molecular energies.} 
In quantum chemistry the performance of approximate QM solvers is usually assessed by
single point calculations at reference 
geometries.~\cite{YuZhaBerHeTru-PCCP-15,MarHea-MP-17,GoeHanBauEhrNajGri-PCCP-17,Gou-PCCP-18}
Now we show the importance of GEO in such comparisons using the S66x8 data set.

In Figure~\ref{fig:s66}, we show plots
comparing GEO
with the total $\left| \Delta E \right|$ errors of MP2 (with a complete basis set) and the PBE0 hybrid with the D3 correction for
the S66 binding energies (other methods are in SI in Figs.~S39-S41). 
Vertical and horizontal dashed lines represent the 
averaged GEO and $\left| \Delta E \right|$, respectively, separated into H-bonds, disperson dominated interactions,
and others.  Both methods have overall mean average energy errors of about 0.5 kcal/mol, but MP2 is worst
for dispersion complexes, while PBE0[D3] is worst for H-bonds.   But clearly, MP2 yields much worse
geometries with GEO values about 4 times larger.  Also, for MP2, there is a strong correlation
between GEO and errors in binding energies.  But for PBE0[D3], GEO errors do not correlate with such errors.  For the geometry of weak bonded complexes, PBE0[D3] is clearly superior to MP2,
but not (overall) for binding energies.

Finally, we consider the effect of varying geometries on binding energy errors in Table 1.
Each row is a method for finding a geometry, each column is the method used for finding the energy,
all averaged over the 66 complexes.  The blue diagonals are the errors of  each method at its own geometry.
The red column at the right is the errors on accurate geometries, given by CCSD(T), i.e., the
best estimate of the exact geometry here.  The numbers are very similar in all cases, suggesting
that improvements in geometry do not matter.  However, this makes the B2PLYP column even more
surprising:  For all methods, the errors are smaller at the B2PLYP geometries than at their own geometries,
sometimes by
almost a factor of 2, and always better than on the accurate geometries!   How can this be?
We explored and found that most approximations overbind the S66 complexes (particularly those bonded by dispersion), and B2LYP typically overestimates the bond lengths. For this reason, energies of approximate methods at the B2PLYP minimum are more accurate than at their own minimum (see the benzene-uracil binding curves shown in Fig.~S42.)

Here we have covered only the most obvious topics
that the GEO concept brings into focus. For main group chemistry and
weak interactions, GEO calculations and analysis yield an ideal tool for
understanding geometric errors and for ranking different approximations, one
that is very different from tables of errors in atomization/binding energies, and could easily be applied to transition state geometries. 
GEO should also be useful for
the lattice parameters of solids, or for geometries of transition metal
compounds, both of which 
require an accurate reference geometry.
Here we analyse errors in geometries obtained from electronic structure methods, but the very same tools are also applicable to molecular mechanics methods.

{\bf Methods.}
Now we go back to the top panel of Figure~\ref{fig:f}a , where we calculate mean GEO for a dataset of 14 small organic molecules, which is the {\em AV5Z} subset of the W4-11-GEOM set produced by Karton and co-workers~\cite{SpaJayKar-JCP-16}. The geometries from this dataset have been optimized at the CCSD(T) level, which is the level of theory that we use for this dataset as a reference. GEOs  for approximate methods that we consider here are obtained as follows. 
The $\tE \left (\G_0\right)$ quantity is obtained from the total energies of each of the approximate method at CCSD(T) geometries. Then we relax CCSD(T) geometries by using each of the approximate methods and this allows us to calculate the $\tE \left (\tG\right)$ term. Finally, we obtain the total energies of each of the approximate method at CCSD(T) geometries to compute $E_0 \left (\tG\right)$ . These energies are obtained from single point CCSD(T)/A'V5Z calculations on the $\tG$ geometries. In this way, we ensure that the level of theory used for the reference single point energy calculations matches the level of theory used for obtaining reference geometries by Karton and co-workers~\cite{SpaJayKar-JCP-16,DunPetWil-JCP-01}. Other GEOs in the paper are computed in the same manner and in SI we provide further computational details.

{\bf Supplementary Information and Data availability.}
The pdf document with Supplementary Information is available at:
\burl{https://dft.uci.edu/pubs/VB20s.pdf}

The authors declare that the data supporting the findings of this study are
available within the paper and its supplementary information files.

{\bf Acknowledgments.}
We thank J. Rez{\'a}c for providing us the S66x8 binding energies for a selection of wavefunction methods. SV acknowledges funding from the Rubicon project (019.181EN.026), which is financed by the Netherlands Organisation for Scientific Research (NWO). KB acknowledges funding from NSF (CHE 1856165). 

{\bf Competing interests.}
The authors declare that there are no competing interests.

{\bf Author contributions.}
SV proposed the idea, KB proposed the analysis, and SV performed all calculations.
Both authors designed the research, analysed the data and wrote the manusript.

\bibliographystyle{naturemag}
\bibliography{all}

\end{document}


\maketitle

\tableofcontents
\newpage

\sf

\listoftables
\listoffigures
\clearpage



\section{Additional GEO mathematical details}

\begin{figure}[htb]
\centering
\includegraphics[width=8cm]{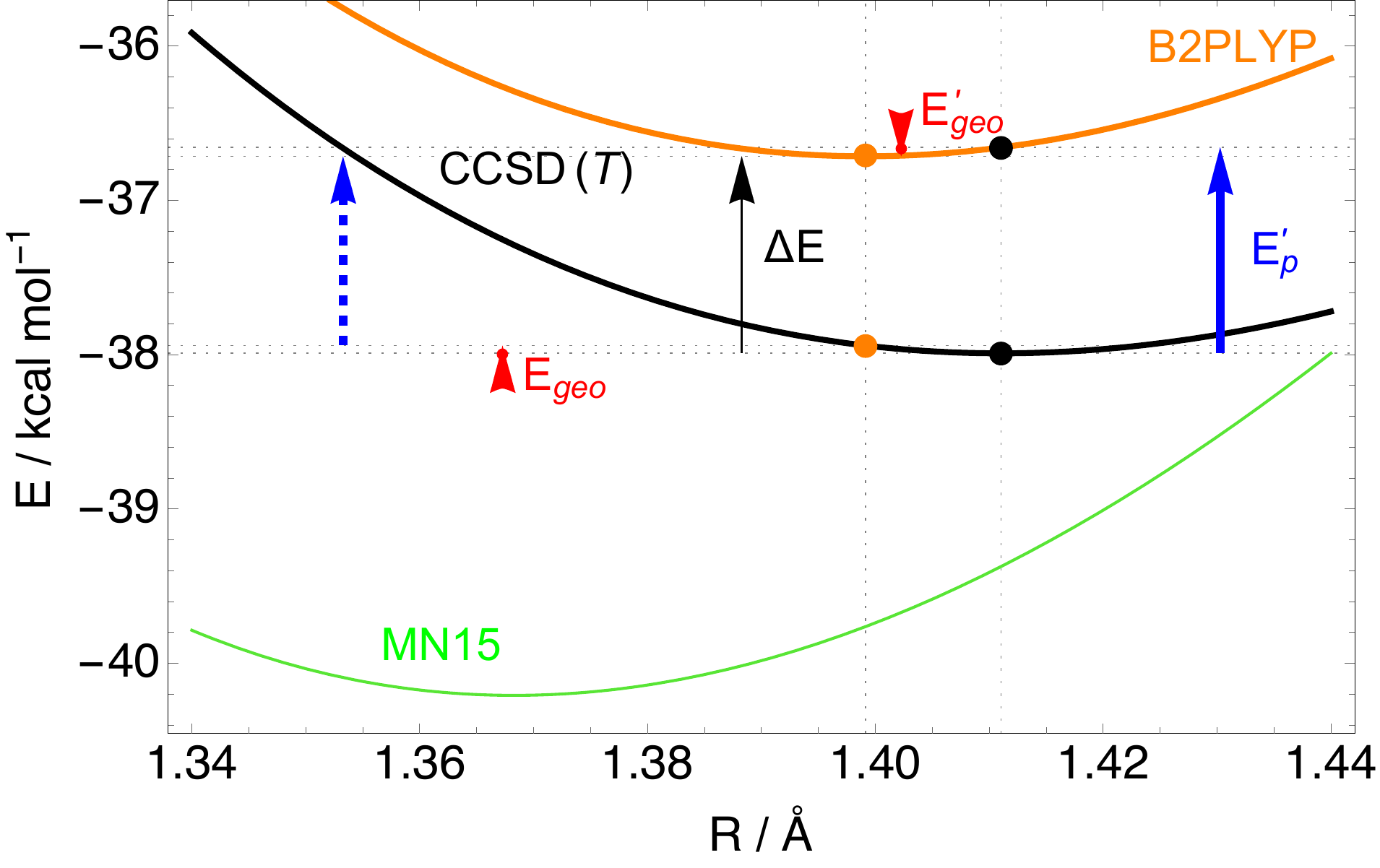}
\includegraphics[width=8cm]{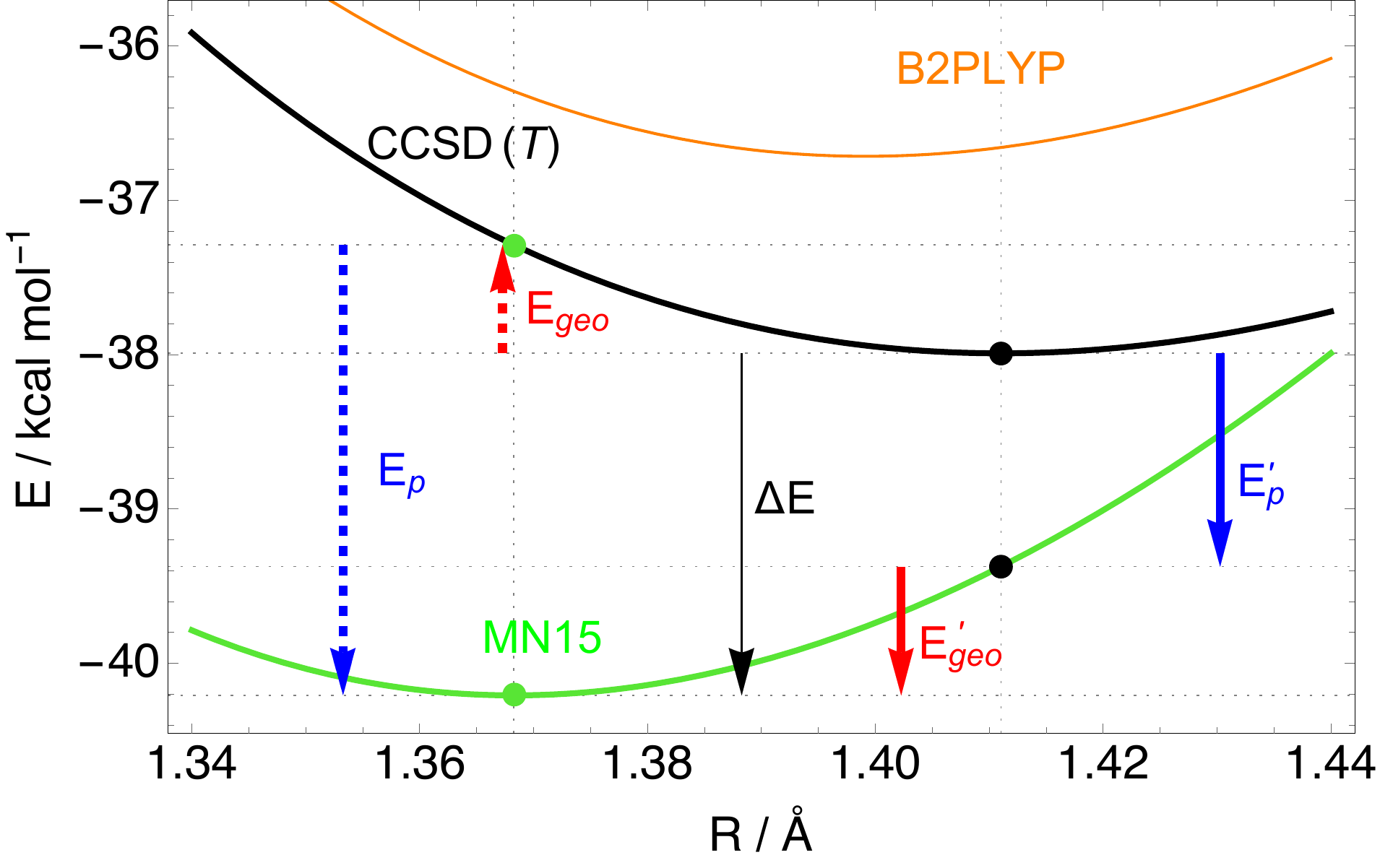}
\caption{The decomposition of $\Delta E$ error by Eqs.~\ref{eq:errold} and~\ref{eq:err2old}) for the atomization energy of the F$_2$ molecules obtained from the B2PLYP ($\geo $ and $\geop$ are negligible ) and MN15 functionals  ($\geo $ and $\geop$ are non-negligible ) }
\label{fig:typ}
\end{figure}

For an approximate quantum-mechanical (QM) solver $\tE \left( \mathbf{G} \right)$, the total error is given by:
\beq \label{eq:errold}
\dE = \tE \left (\tG\right) - E\left (\G_0 \right),
\eeq
where $E \left ( \bf{G} \right)$ is the exact ground-state energy at geometry $\G$, $\G_0$ is the exact and $\tG$ is the approximate geometry. Thus, $\dE$ contains errors both due to the approximate geometry and approximate energy.  To decompose this error into the GEO and "non-GEO" part, we add and subtract $E \left( \tG\right)$ to the r.h.s of Eq.~\ref{eq:errold}:
\beq \label{eq:err1old}
\dE = \underbrace{E \left (\tG\right )  - E\left (\G_0\right )}_{\geo \geq 0}
+
\underbrace{\tE\left (\tG\right )-E\left (\tG\right )}_{E_\p}.
\eeq
We can also add and subtract $\tE \left( \G_0\right)$ to the r.h.s of Eq.~\ref{eq:errold}, to obtain an alternative form of Eq.~\ref{eq:err1old}:
\beq \label{eq:err2old}
\dE = \underbrace{\tE \left (\tG\right)  - \tE \left(\G\right)}_{\geop \leq 0}
+ 
\underbrace{\tE \left(\G_0\right)-E \left(\G_0\right)}_{E_\p^\prime}.
\eeq
The signs of $\dE_\g$ and $\dtE_\g$  also dictate the following chain of inequalities:
\beq
E_\p \leq \Delta E \leq E_\p^\prime . 
\eeq
For an illustration of the error decomposition by means of Eqs.~\ref{eq:errold} and~\ref{eq:err2old}, see Fig.~\ref{fig:typ}.  As discussed throughout the text, $\geo$ is typically accurately approximated by $- \geop$, and thus : $\geo \approx \frac{1}{2} \left( E_\p^\prime - E_p \right)$ (see Fig.~\ref{fig:typ}).

For equilibrium structures (minima of potential energy surfaces), $\geo$ is always positive. On the other hand, in the case of transition states (the first order saddle points of potential energy surfaces), the sign of $\geo$  is not definite. In that case, the $\mathbf{H}_0$ Hessian will have one negative eigenvalue. Thus, one term on the r.h.s of Eq~\ref{eq:gaNORM} of the underlying $\geoh$ is negative, and all other are positive.

\FloatBarrier

\section{Additional details on the results in Figure~\ref{fig:f}a (top panel)}
The CCSD(T)/A'V5Z (see Ref.~\cite{SpaJayKar-JCP-16}
for the basis set description) $E\left (\G\right)$ values and $\G$ geometries have been taken from 
Ref.~\cite{SpaJayKar-JCP-16}. All other results have been obtained from the Gaussian16 calculations by always employing the same (A'V5Z) basis set.
The $\geop$ and $\geo$ errors for individual molecules and employed approximate methods are given in Tables~\ref{tab_de} and~\ref{tab_dbe}, respectively. 
In Figure~\ref{fig_ma}, we show the MAE for the atomization energies for the same set of molecules.
The plot In Figure~\ref{fig_ebars} shows the data of Table~\ref{tab_dbe}.

\begin{table*}[!h]
\centering
\footnotesize
\begin{tabular}{llllllllllllll}
      & HF    & PBE   & BLYP  & TPSS  & B3LYP & MN15  & $\omega$B97 & $\omega$B97XD & LSDA  & PBE {[}D3{]} & MP2   & B2PLYP & PBE0  \\ \hline \hline
acetaldehyde & 0.957 & 0.203 & 0.203 & 0.067 & 0.049 & 0.067 & 0.039      & 0.112        & 0.412 & 0.207        & 0.032 & 0.018  & 0.122 \\
acetic       & 1.431 & 0.322 & 0.567 & 0.227 & 0.038 & 0.093 & 0.024      & 0.091        & 0.375 & 0.324        & 0.030 & 0.016  & 0.077 \\
c2h3f        & 0.698 & 0.165 & 0.292 & 0.109 & 0.059 & 0.084 & 0.072      & 0.107        & 0.263 & 0.169        & 0.024 & 0.037  & 0.094 \\
c2h5f        & 0.391 & 0.277 & 0.495 & 0.192 & 0.036 & 0.045 & 0.023      & 0.017        & 0.331 & 0.287        & 0.033 & 0.024  & 0.035 \\
ch2ch        & 0.191 & 0.150 & 0.118 & 0.042 & 0.085 & 0.092 & 0.063      & 0.083        & 0.402 & 0.152        & 0.866 & 0.058  & 0.074 \\
ch2nh2       & 0.276 & 0.166 & 0.182 & 0.073 & 0.056 & 0.223 & 0.040      & 0.053        & 0.577 & 0.171        & 0.054 & 0.036  & 0.065 \\
ch3nh        & 0.284 & 0.228 & 0.168 & 0.064 & 0.029 & 0.074 & 0.030      & 0.049        & 0.829 & 0.234        & 0.041 & 0.022  & 0.074 \\
ch3nh2       & 0.384 & 0.188 & 0.225 & 0.080 & 0.035 & 0.116 & 0.039      & 0.042        & 0.411 & 0.196        & 0.030 & 0.025  & 0.058 \\
ethanol      & 0.622 & 0.270 & 0.440 & 0.168 & 0.036 & 0.084 & 0.019      & 0.049        & 0.445 & 0.278        & 0.036 & 0.020  & 0.056 \\
formic       & 1.448 & 0.265 & 0.428 & 0.174 & 0.038 & 0.084 & 0.029      & 0.099        & 0.269 & 0.278        & 0.009 & 0.009  & 0.088 \\
methanol     & 0.555 & 0.207 & 0.311 & 0.126 & 0.018 & 0.081 & 0.017      & 0.043        & 0.319 & 0.221        & 0.016 & 0.011  & 0.050 \\
propane      & 0.212 & 0.225 & 0.282 & 0.085 & 0.034 & 0.039 & 0.009      & 0.016        & 0.432 & 0.225        & 0.058 & 0.027  & 0.038 \\
propene      & 0.468 & 0.193 & 0.156 & 0.055 & 0.076 & 0.082 & 0.074      & 0.115        & 0.461 & 0.195        & 0.048 & 0.048  & 0.109 \\
propyne      & 0.991 & 0.171 & 0.090 & 0.045 & 0.138 & 0.135 & 0.136      & 0.200        & 0.499 & 0.173        & 0.047 & 0.051  & 0.170 \\ \hline
mean         & 0.636 & 0.216 & 0.283 & 0.108 & 0.052 & 0.093 & 0.044      & 0.077        & 0.430 & 0.222        & 0.094 & 0.029  & 0.079 \\
RMS          & 0.755 & 0.222 & 0.317 & 0.122 & 0.060 & 0.102 & 0.054      & 0.090        & 0.452 & 0.228        & 0.234 & 0.032  & 0.087
\end{tabular}
\caption{$\geo$ values of Figure~\ref{fig:f}a (top panel) in kcal/mol for individual molecules. }
\label{tab_dbe}
\end{table*}
\FloatBarrier

\begin{table*}[!h]
\centering
\footnotesize
\begin{tabular}{llllllllllllll}
         & HF     & PBE    & BLYP   & TPSS   & B3LYP  & MN15   & ωB97   & ωB97XD & LSDA   & PBE {[}D3{]} & MP2    & B2PLYP & PBE0   \\ \hline \hline 
acetaldehyde & -1.055 & -0.196 & -0.196 & -0.065 & -0.050 & -0.071 & -0.042 & -0.117 & -0.393 & -0.199       & -0.032 & -0.018 & -0.126 \\
acetic       & -1.583 & -0.311 & -0.539 & -0.219 & -0.037 & -0.095 & -0.025 & -0.094 & -0.361 & -0.314       & -0.030 & -0.016 & -0.080 \\
c2h3f        & -0.761 & -0.160 & -0.274 & -0.104 & -0.058 & -0.087 & -0.075 & -0.110 & -0.260 & -0.164       & -0.024 & -0.037 & -0.097 \\
c2h5f        & -0.419 & -0.267 & -0.462 & -0.182 & -0.034 & -0.048 & -0.023 & -0.017 & -0.319 & -0.277       & -0.034 & -0.023 & -0.035 \\
ch2ch        & -0.192 & -0.157 & -0.124 & -0.057 & -0.127 & -0.137 & -0.105 & -0.133 & -0.443 & -0.159       & -1.224 & -0.100 & -0.118 \\
ch2nh2       & -0.297 & -0.160 & -0.174 & -0.071 & -0.055 & -0.203 & -0.039 & -0.053 & -0.548 & -0.165       & -0.053 & -0.036 & -0.064 \\
ch3nh        & -0.308 & -0.213 & -0.159 & -0.061 & -0.027 & -0.074 & -0.032 & -0.050 & -0.749 & -0.219       & -0.042 & -0.022 & -0.073 \\
ch3nh2       & -0.426 & -0.183 & -0.217 & -0.078 & -0.034 & -0.109 & -0.039 & -0.042 & -0.378 & -0.190       & -0.031 & -0.025 & -0.057 \\
ethanol      & -0.685 & -0.263 & -0.419 & -0.162 & -0.036 & -0.084 & -0.020 & -0.050 & -0.429 & -0.271       & -0.037 & -0.020 & -0.056 \\
formic       & -1.603 & -0.256 & -0.407 & -0.167 & -0.038 & -0.086 & -0.030 & -0.103 & -0.258 & -0.269       & -0.009 & -0.009 & -0.090 \\
methanol     & -0.613 & -0.201 & -0.295 & -0.121 & -0.018 & -0.080 & -0.018 & -0.044 & -0.304 & -0.215       & -0.017 & -0.011 & -0.050 \\
propane      & -0.226 & -0.221 & -0.274 & -0.082 & -0.034 & -0.040 & -0.009 & -0.017 & -0.415 & -0.221       & -0.059 & -0.027 & -0.038 \\
propene      & -0.511 & -0.187 & -0.152 & -0.054 & -0.077 & -0.085 & -0.079 & -0.119 & -0.448 & -0.190       & -0.048 & -0.052 & -0.112 \\
propyne      & -1.093 & -0.164 & -0.086 & -0.044 & -0.141 & -0.139 & -0.145 & -0.208 & -0.484 & -0.165       & -0.046 & -0.052 & -0.175 \\ \hline 
mean         & -0.698 & -0.210 & -0.270 & -0.105 & -0.055 & -0.095 & -0.049 & -0.083 & -0.413 & -0.216       & -0.120 & -0.032 & -0.084 \\
RMS          & ~0.832  & ~0.215  & ~0.301  & ~0.118  & ~0.065  & ~0.104  & ~0.061  & ~0.097  & ~0.431  & ~0.221        & ~0.329  & ~0.039  & ~0.092 
\end{tabular}
\caption{$\geop$ values of Figure~\ref{fig:f}a (top panel) in kcal/mol for individual molecules. }
\label{tab_de}
\end{table*}
\FloatBarrier

\begin{figure}
\centering
\includegraphics[width=10cm]{fsup/mean_atom.pdf}
\includegraphics[width=10cm]{fsup/corr.pdf}
\captionof{figure}[GEO error vs. atomization energy errors for the set of molecules and approximations shown in Fig.~\ref{fig:f}a (top panel)]{Top panel: MAEs for the atomization energies for the set of molecules (listed in Tables~\ref{tab_de} and~\ref{tab_dbe}) [Fig.~\ref{fig:f}a (top panel)] obtained from different methods. The LSDA and HF MAEs are divided by the factor of $4.5$. Bottom panel: MAEs from the top panel vs. average $\geo$ errors for the same set of molecules.}
\label{fig_ma}
\end{figure}
\FloatBarrier

\begin{figure}
\centering
\includegraphics[width=10cm]{fsup/ginv.pdf}
\captionof{figure}[GEO error vs. atomization energy errors for the set of molecules and approximations shown in Fig.~\ref{fig:f}a (top panel)]{The {\em inverted} $\gamma$, $\gamma_{inv}=\sqrt{\frac{2 \geo}{\K}}$, averaged over the molecular dataset considered in ~\ref{fig:f}a (top panel). The underlying $\geo$ values are given in Table~\ref{tab_dbe}, and the $\K$ values are given in Table~\ref{tab:k1}. }
\label{fig_ginv}
\end{figure}
\FloatBarrier

\begin{figure}
\centering
\includegraphics[width=13cm]{fsup/ebars14.pdf}
\captionof{figure}[Plots showing data of Table~\ref{tab_dbe}. ]{Plots showing data of Table~\ref{tab_dbe}. The order of molecules is the same as in Table~\ref{tab_dbe}.}
\label{fig_ebars}
\end{figure}
\FloatBarrier

\section{Additional details on the results in Figure~\ref{fig:f}a (bottom panel)}

The dataset of molecules used in Figure~\ref{fig:f}a (bottom panel) is shown in Figure~\ref{fig_grid}. 
The PBEh-3c and HF-3c calculations have been performed with the ORCA package\cite{ORCA4}.
The GFN1-xTB and DFTB3 [D3] calculations have been performed with the ADF package\cite{ADF}. 
All other calculations have been performed with the G16 package. 
We used aug-cc-pVTZ\cite{Dun-JCP-89}  basis set for all DFT and HF calculations, except for PBEh-3c and HF-3c, which come with their own basis set.
For DFTB3 [D3] calculations, we have used the set of parameters of Ref.~\cite{DFTB3} in tandem with the D3 correction with the Becke-Johnson damping function\cite{BJ} .
The set of parameters for PM6 and PM7 are those implemented in G16 when the keywords {\em "PM6"} and {\em "PM7"}, respectively, are used. 
The $\geo$ errors for individual molecules (Figure~\ref{fig_grid}) and employed approximate methods are shown in Figure~\ref{fig_mol}, and the underlying data values are given in Table~S4. 
\FloatBarrier

\renewcommand{\thesubfigure}{\arabic{subfigure}}
\begin{figure*}
\centering
\begin{tabular}{cccc}
\subcaptionbox{\label{1}}{\includegraphics[width=3cm]{fsup/f/E13.pdf}} &
\subcaptionbox{\label{2}}{\includegraphics[width=3cm]{fsup/f/E20.pdf}} &
\subcaptionbox{\label{3}}{\includegraphics[width=3cm]{fsup/f/E22.pdf}} &
\subcaptionbox{\label{4}}{\includegraphics[width=3cm]{fsup/f/E23.pdf}}\\
\subcaptionbox{\label{1}}{\includegraphics[width=3cm]{fsup/f/E32.pdf}} &
\subcaptionbox{\label{2}}{\includegraphics[width=3cm]{fsup/f/E34.pdf}} &
\subcaptionbox{\label{3}}{\includegraphics[width=3cm]{fsup/f/P13.pdf}} &
\subcaptionbox{\label{4}}{\includegraphics[width=3cm]{fsup/f/P20.pdf}}\\
\subcaptionbox{\label{1}}{\includegraphics[width=3cm]{fsup/f/P23.pdf}} &
\subcaptionbox{\label{2}}{\includegraphics[width=3cm]{fsup/f/P32.pdf}} &
\subcaptionbox{\label{3}}{\includegraphics[width=3cm]{fsup/f/P34.pdf}} &
\subcaptionbox{\label{4}}{\includegraphics[width=3cm]{fsup/f/rezorcinol.pdf}}\\
\subcaptionbox{\label{1}}{\includegraphics[width=3cm]{fsup/f/4-nitrophenol.pdf}} &
\subcaptionbox{\label{2}}{\includegraphics[width=3cm]{fsup/f/naphthalene.pdf}} &
\subcaptionbox{\label{3}}{\includegraphics[width=3cm]{fsup/f/napht_oh_nh2.pdf}} &
\subcaptionbox{\label{4}}{\includegraphics[width=3cm]{fsup/f/naphtol.pdf}}\\
\end{tabular}
\caption{Molecules used in error plots of Figure~\ref{fig:f}a (bottom panel).}
\label{fig_grid}
\end{figure*}
\FloatBarrier

\begin{figure*}
\centering
\includegraphics[width=8.5cm]{fsup/med1.pdf}
\includegraphics[width=8.5cm]{fsup/med2.pdf}
\centering
\includegraphics[width=8.5cm]{fsup/med3.pdf}
\includegraphics[width=8.5cm]{fsup/med4.pdf}
\captionof{figure}[$\geo$ errors for different approximate methods for the set of molecules shown in Figure~\ref{fig_grid}.]{Plots showing $\geo$ errors for different approximate methods for the set of molecules shown in Figure~\ref{fig_grid}.
Averaged error bars are shown in Figure~\ref{fig:f}a (bottom panel).  Note the difference in the y-axes scales.}
\label{fig_mol}
\end{figure*}
\FloatBarrier

\begin{sidewaystable*}[]
\tiny
\centering
\begin{tabular}{lllllllllllllllllll}
      & $\omega$ B97 & B3LYP & $\omega$B97XD & MN15  & MP2   & TPSS  & PBE   & PBE {[}D3{]} & BLYP  & LSDA  & HF    & DFTB3 {[}D3{]} & GFN1-xTB & HF-3c & PBEh-3c & PM6   & PM7   & PBE0  \\ \hline \hline
1    & 0.144        & 0.018 & 0.138         & 0.063 & 0.054 & 0.186 & 0.472 & 0.473        & 0.557 & 0.858 & 0.763 & 0.266          & 0.155    & 0.521 & 0.105   & 0.806 & 0.749 & 0.104 \\
2    & 0.081        & 0.021 & 0.142         & 0.082 & 0.064 & 0.154 & 0.330 & 0.338        & 0.459 & 0.532 & 1.222 & 0.812          & 0.142    & 0.269 & 0.292   & 2.332 & 1.327 & 0.154 \\
3    & 0.052        & 0.017 & 0.094         & 0.104 & 0.050 & 0.175 & 0.407 & 0.410        & 0.541 & 0.882 & 0.677 & 0.424          & 0.245    & 1.090 & 0.154   & 0.808 & 0.747 & 0.100 \\
4    & 0.055        & 0.052 & 0.032         & 0.363 & 0.107 & 0.317 & 0.561 & 0.556        & 0.940 & 1.511 & 0.480 & 0.552          & 1.028    & 1.370 & 0.133   & 1.908 & 2.059 & 0.115 \\
5    & 0.064        & 0.043 & 0.180         & 0.283 & 0.081 & 0.327 & 0.520 & 0.514        & 0.980 & 1.050 & 1.559 & 1.042          & 0.844    & 1.823 & 0.461   & 2.856 & 2.788 & 0.213 \\
6    & 0.056        & 0.014 & 0.128         & 0.107 & 0.044 & 0.243 & 0.484 & 0.477        & 0.703 & 1.153 & 1.004 & 0.474          & 0.258    & 0.390 & 0.253   & 1.381 & 1.158 & 0.143 \\
7    & 0.329        & 0.080 & 0.196         & 0.201 & 0.316 & 0.210 & 0.512 & 0.497        & 0.646 & 0.977 & 1.232 & 0.423          & 0.708    & 2.694 & 0.179   & 2.966 & 3.893 & 0.120 \\
8    & 0.161        & 0.028 & 0.192         & 0.129 & 0.074 & 0.155 & 0.316 & 0.326        & 0.498 & 0.548 & 1.616 & 0.600          & 0.292    & 0.231 & 0.464   & 2.560 & 1.825 & 0.214 \\
9    & 0.060        & 0.054 & 0.041         & 0.331 & 0.112 & 0.312 & 0.558 & 0.554        & 0.949 & 1.523 & 0.440 & 0.676          & 1.204    & 1.360 & 0.167   & 2.251 & 1.966 & 0.134 \\
10   & 0.089        & 0.075 & 0.157         & 0.199 & 0.156 & 0.260 & 0.442 & 0.439        & 0.739 & 1.397 & 1.704 & 0.615          & 0.794    & 2.022 & 0.434   & 2.244 & 2.328 & 0.171 \\
11   & 0.134        & 0.019 & 0.181         & 0.232 & 0.056 & 0.258 & 0.496 & 0.502        & 0.770 & 1.283 & 1.059 & 0.906          & 0.645    & 0.750 & 0.410   & 1.481 & 1.598 & 0.194 \\
12   & 0.049        & 0.010 & 0.164         & 0.114 & 0.040 & 0.267 & 0.447 & 0.444        & 0.777 & 0.833 & 1.479 & 0.978          & 0.342    & 0.393 & 0.383   & 2.164 & 1.931 & 0.157 \\
13   & 0.440        & 0.055 & 0.492         & 0.323 & 0.064 & 0.325 & 0.489 & 0.491        & 1.119 & 1.058 & 4.191 & 2.964          & 1.385    & 0.740 & 1.165   & 2.081 & 1.802 & 0.497 \\
14   & 0.235        & 0.012 & 0.193         & 0.091 & 0.070 & 0.207 & 0.499 & 0.497        & 0.681 & 0.859 & 1.095 & 0.304          & 0.097    & 1.005 & 0.158   & 0.775 & 0.400 & 0.129 \\
15   & 0.244        & 0.027 & 0.253         & 0.212 & 0.122 & 0.330 & 0.643 & 0.636        & 1.041 & 1.651 & 2.478 & 0.984          & 0.409    & 1.860 & 0.752   & 2.515 & 1.497 & 0.200 \\
16   & 0.245        & 0.015 & 0.241         & 0.142 & 0.083 & 0.287 & 0.573 & 0.566        & 0.874 & 1.239 & 1.651 & 0.653          & 0.279    & 1.141 & 0.323   & 1.866 & 1.356 & 0.182 \\ \hline
mean & 0.152        & 0.034 & 0.176         & 0.186 & 0.093 & 0.251 & 0.484 & 0.483        & 0.767 & 1.085 & 1.416 & 0.792          & 0.552    & 1.104 & 0.365   & 1.937 & 1.714 & 0.177 \\
RMS  & 0.190        & 0.040 & 0.203         & 0.208 & 0.114 & 0.258 & 0.491 & 0.489        & 0.792 & 1.132 & 1.665 & 0.998          & 0.676    & 1.303 & 0.450   & 2.054 & 1.900 & 0.198
\end{tabular}
\label{tab_m}
\captionof{table}[$\geo$ values Figure~\ref{fig:f}a (bottom panel) for individual molecules and approximate methods.]{$\geo$ values Figure~\ref{fig:f}a (bottom panel) for individual molecules and approximate methods, all values are in kcal/mol. The molecules are ordered in Figure~\ref{fig_grid}.}
\end{sidewaystable*}
\FloatBarrier

\section{Details on the results of Figure~\ref{fig:f}b}
\FloatBarrier

\begin{figure*}
\includegraphics[width=1.0\columnwidth]{fsup/g/geok.pdf}
\captionof{figure}[$\geo$ and $\geod$ for a set of molecules shown in Fig.~\ref{fig:f}b.]{Plots comparing $\geo$ and $\geod$ for selected methods and for a set of molecules shown in Fig.~\ref{fig:f}b.}
\label{fig:geok2}
\end{figure*}

\begin{figure*}
\includegraphics[width=1.0\columnwidth]{fsup/g/ab.pdf}
\captionof{figure}[Bond vs. angle contribution to $\geos$ for a set of molecules shown in Fig.~\ref{fig:f}b.]{Plots comparing total bond length and angle errors contribution to $\geos$ for selected methods and for a set of molecules shown in Fig.~\ref{fig:f}b.}
\label{fig:allab}
\end{figure*}

\begin{figure*}
\includegraphics[width=1.0\columnwidth]{fsup/g/b2.pdf}
\captionof{figure}[Total bond vs. double bond contribution to $\geos$ for a set of molecules shown in Fig.~\ref{fig:f}b.]{Plots comparing total bond and double bond length errors contribution to $\geos$ for selected methods and for a set of molecules shown in Fig.~\ref{fig:f}b.}
\label{fig:allb2}
\end{figure*}

\begin{figure*}
\includegraphics[width=1.0\columnwidth]{fsup/g/ghs.pdf}
\captionof{figure}[$\geo$, $\geoh$ and $\geos$ plots for a set of molecules shown in Fig.~\ref{fig:f}b.]{Plots comparing $\geo$, $\geoh$ and $\geos$ for selected methods and for a set of molecules shown in Fig.~\ref{fig:f}(b).}
\label{fig:allghs}
\end{figure*}

\begin{figure}
\includegraphics[width=0.5\columnwidth]{fsup/g/1.pdf}
\includegraphics[width=0.5\columnwidth]{fsup/g/2.pdf}
\captionof{figure}[PBE and B3LYP errors in bond lengths for a set of molecules shown in Fig.~\ref{fig:f}b.]{PBE and B3LYP errors in bond lengths for a set of molecules shown in Fig.~\ref{fig:f}b. Left panel: relative errors; middle panel: mean absolute errors.}
\label{fig:aller}
\end{figure}
\FloatBarrier

\begin{figure}
\includegraphics[width=0.3\columnwidth]{fsup/g/singles.pdf}
\includegraphics[width=0.3\columnwidth]{fsup/g/doubles.pdf}
\includegraphics[width=0.3\columnwidth]{fsup/g/angs.pdf}
\captionof{figure}[Single bond, double bond, angles $\geos$ rankings for a set of molecules shown in Fig.~\ref{fig:f}b.]{Rankings of approximations for a set of molecules shown in Fig.~\ref{fig:f}b based on averaged-$\geos$ (over a number of molecules) contributions from: single bonds (left panel), double bonds (middle panel), angles (right panel).}
\label{fig:simpleD}
\end{figure}
\FloatBarrier

\clearpage

tube
\section{The GEO analysis for the water molecule} 
\label{sec_water}

\begin{figure}
\centering
\includegraphics[width=9cm]{fsup/w1.pdf}
\captionof{figure}[Bond length, bond angle, and the $\geo$ errors for the water molecule]{The error in the bond length, bond angle, and the $\geo$ errors for the water molecule obtained from various approximations.}
\label{fig:w1}
\end{figure}

We use CCSD(T)/A'V6Z~\cite{SpaJayKar-JCP-16} as a reference method in all calculations performed in this section. The same basis set is employed in calculations involving approximate electronic structure methods and all calculations have been performed with the Gaussian16 package. 

To show how GEO depends on errors in individual bond lengths and angles,
we consider the water molecule in internal coordinates. 
It has three degrees of freedom, but to analyse the $\geo$ error we need just two:  $a$ and $\theta$
(see Figure~\ref{fig:w1}) since the lengths of the two \ce{O-H} bonds are the same  within any approximation. In Figure~\ref{fig:w1}, we also 
show the errors of different methods in the bond length, angle, together with the $\geo$ errors. 

\begin{figure}
\centering
\includegraphics[width=7cm]{fsup/w2.pdf} 
\includegraphics[width=7cm]{fsup/w3.pdf} \\
\includegraphics[width=7cm]{fsup/w4.pdf} 
\includegraphics[width=7cm]{fsup/w5.pdf} 
\captionof{figure}[Different $\geo$ vs. $\geoh$ vs $\geos$ plots for the water molecule.]{ (a) Contour plot of GEO as a function of errors in the bond lengths for the water molecule (also shown in the top panel of Fig.~\ref{fig:f}a);
(b) same plot from the harmonic approximation (Eq.~\ref{eq:q});  (c) same plot from the {\em simple} quadratic approximation (Eq.~\ref{eq:qs});
(d) Comparison of the $\geo$ , $\geoh$, and $\geos$ obtained from different methods. }
\label{fig:w2}
\end{figure}

From a series of CCSD(T) calculations at fixed geometries, we can see how 
$\geo$ depends on $\Delta a$ and $\Delta \theta$, the error in the bond length and bond angle, respectively. 
The resulting contour plot is shown in Figure~\ref{fig:w2}a. From this plot, we can see that the $\geo$  error increases
much more quickly with the error in the bond length than in the bond angle.
To better understand this observation, we consider the harmonic approximation to $\geo$ (Eq.~\ref{eq:ga}). For clarity,
we write this equation again:
\beq \label{eq:gaa}
\geo \approx \geoh= \half  \Delta \tilde{\G}^\intercal \H \left( \G \right)  \Delta \tilde{\G}.
\eeq
For the water molecule in internal coordinates, we can now write $\geoh$ as a function of:
\beq \label{eq:wg}
\Delta \tilde{\G}= \left( 2 \Delta a,  \Delta  \theta    \right). 
\eeq
The underlying $\H \left( \G \right) $ matrix in internal coordinates is given by:
\beq \label{eq:wh}
 \H \left( \G \right) = \begin{pmatrix} 
k_{aa} & k_{a \theta} \\
k_{ \theta a}&k_{ \theta  \theta} 
\end{pmatrix}
\sim
\begin{pmatrix} 
0.122 {\rm pm}^{-2} & 0.006^{\circ^{-1}}{\rm pm}^{-1} \\
0.006^{\circ^{-1}}{\rm pm}^{-1} & 0.031$^{\circ^{-2}}$
\end{pmatrix}  {\rm kcal/mol}
\eeq
where $k$-s are the force constants at the highly accurate reference [CCSD(T)] geometry. 
Plugging Eqs.~\ref{eq:wg} and~\ref{eq:wh} into Eq.~\ref{eq:gaa}, we obtain:
\beq \label{eq:q}
\geoh \left ( \Delta a, \Delta \theta \right ) =  k_{aa} \Delta a^2 + \half k_{\theta \theta} \Delta \theta^2 
+2 k_{a \theta} \Delta a \Delta \theta,
\eeq
where we also used the Hessian symmetry: $ k_{a \theta}=k_{\theta a}$. In Eq.~\ref{eq:wh}, the units were chosen such that 
$\Delta a$ errors in pm and $\Delta \tilde  \theta$ errors in degrees yield $\geoh}$ in kcal/mol.
In Figure~\ref{fig:w2}b, we show the $\geoh$ contour plot as a function of $\Delta a$ and $\Delta \tilde  \theta$.
We can see that it is in a very close agreement with its {\em exact} counterpart (Figure~\ref{fig:w2}a). 
A further simplification of $\geoh$ (Eq.~\ref{eq:q}) obtained by setting $k_{a \theta}$ to $0$:
\beq  \label{eq:qs}
\geoh \left ( \Delta a, \Delta \theta \right )  \approx \geos \left ( \Delta a, \Delta \theta \right )= k_{aa} \Delta a^2 + \half k_{\theta \theta} \Delta \theta^2, 
\eeq
gives a sensible approximation to $\dE_\g \left (\Delta a, \Delta \theta \right )$, as it can be seen from Figure~\ref{fig:w2}c. 
In Figure~\ref{fig:w2}d, we compare the size of the three errors  from different methods and we can see that 
both $\geoh$ and $\geos}$ are in a close agreement with $\geo$.
Thus, both Eq.~\ref{eq:q} and~\ref{eq:qs} give us a simple yet very accurate representation of the GEO for the water molecule. 
From Eq.~\ref{eq:q} and values of the force constants (Eq.~\ref{eq:wh}), we can see that an error of 2pm in the bond length results in the GEO of $\sim$0.5kcal/mol. On the other hand, because of the smaller size of 
 $k_{\theta \theta}$, we can see that the GEO for the water molecule is much more forgiving of the bond angle error, since an error of 2$^{\circ}$, results in the GEO of only $\sim$0.06kcal/mol. Interestingly, Eq.~\ref{eq:qs} can be rearranged to be in a form of an ellipse equation:
 \beq  \label{eq:qe}
 \frac{k_{aa}}{\geoh} \Delta a^2 + \frac{1}{2 \geoh} k_{\theta \theta} \Delta \theta^2 
+2 \frac{k_{a \theta}}{ \geoh} \Delta a \Delta \theta = 1.
 \eeq
 \begin{figure}
\centering
\includegraphics[width=7cm]{fsup/ellipse.pdf} 
\caption{ The ellipses satisfying Eqs.~\ref{eq:qe} and~\ref{eq:qs2} for the water molecule.}
\label{fig:el}
\end{figure}
Similarly, rearranging Eq.~\ref{eq:qs}, we obtain an axis-aligned ellipse equation:
\beq  \label{eq:qs2}
\frac{k_{aa} }{\geos} \Delta a^2 + \frac{ k_{\theta \theta}}{2  \geos} \Delta \theta^2= 1.
\eeq

Since $k_{a \theta}$ is small, the two ellipses of Eq.~\ref{eq:qe} and~\ref{eq:qs2} are in a close agreement, as it can be seen from Figure~\ref{fig:el}. 

\section{GEO for small molecules in internal coordinates} 
In this section we show detailed GEO analysis for a set of small molecules.  For all molecules, CCSD(T) has been used as a reference and all calculation have been obtained with the G16 package. The following basis set have ben used: A'V6Z for the water molecule (Figure~\ref{fig:m1}), aug-cc-pVQZ for \ce{H2S} (Figure~\ref{fig:m2}) and ethene (Figure~\ref{fig:m5}), and aug-cc-pVTZ for \ce{H2CO} (Figure~\ref{fig:m4}),  methanol (Figure~\ref{fig:m6}), cyclopropane(Figure~\ref{fig:m6}).
\FloatBarrier
\clearpage

\begin{figure}
\begin{subfigure}{0.45\textwidth}
 \centering
\includegraphics[width=0.65\linewidth]{fsup/h2o/mol.pdf}  
    \caption{}
    \end{subfigure}
    \begin{subfigure}{.45\textwidth}
  \centering
    \includegraphics[width=1\linewidth]{fsup/h2o/eg.pdf}  
     \caption{}
    \end{subfigure}
   \begin{subfigure}{1\textwidth}
    \includegraphics[width=0.5\linewidth]{fsup/h2o/er1.pdf}  
    \includegraphics[width=0.5\linewidth]{fsup/h2o/er2.pdf} 
      \caption{}
     \end{subfigure}
         \begin{subfigure}{0.5\textwidth}
  \centering
                \includegraphics[width=1\linewidth]{fsup/h2o/s.pdf} 
                            \caption{} 
    \end{subfigure}
\begin{subfigure}{0.5\textwidth}
       \centering
            \hspace*{0.5cm}  \includegraphics[width=1\linewidth]{fsup/h2o/w.pdf}  
            \caption{}
\end{subfigure}
\captionof{figure}[GEO analysis for the water molecule in internal coordinates]{The GEO analysis for the water molecule. Panel (a): considered degrees of freedom;
(b):  Comparison of GEO with its $harm$ and $simple$ (labeled as "$s$" in the plot for brevity) approximations obtained from various methods; (c): errors in specific $i$-th degree of freedom specified in (a), with insets showing the resulting $ E_{geo}^{simple,i}$ (see Eq.~\ref{eq:gas}); (d) different $ E_{geo}^{simple,i}$ contributions  to the {\em total} $\geos$; (e) the resulting {\em weights}, i.e. $ E_{geo}^{simple,i}/ \geos$. }
\label{fig:m1}
\end{figure}
\FloatBarrier
\clearpage

\begin{figure}
\begin{subfigure}{0.45\textwidth}
 \centering
\includegraphics[width=0.65\linewidth]{fsup/h2s/mol.pdf}  
    \caption{}
    \end{subfigure}
    \begin{subfigure}{.45\textwidth}
  \centering
    \includegraphics[width=1\linewidth]{fsup/h2s/eg.pdf}  
     \caption{}
    \end{subfigure}
   \begin{subfigure}{1\textwidth}
    \includegraphics[width=0.5\linewidth]{fsup/h2s/er1.pdf}  
    \includegraphics[width=0.5\linewidth]{fsup/h2s/er2.pdf} 
      \caption{}
     \end{subfigure}
         \begin{subfigure}{0.5\textwidth}
  \centering
                \includegraphics[width=1\linewidth]{fsup/h2s/s.pdf} 
                            \caption{} 
    \end{subfigure}
\begin{subfigure}{0.5\textwidth}
       \centering
            \hspace*{0.5cm}  \includegraphics[width=1\linewidth]{fsup/h2s/w.pdf}  
            \caption{}
\end{subfigure}
\caption{The same plots as in Figure~\ref{fig:m1}, but for the \ce{H2S} molecule.}
\label{fig:m2}
\end{figure}
\FloatBarrier
\clearpage

\begin{figure}
\begin{subfigure}{0.45\textwidth}
 \centering
\includegraphics[width=0.65\linewidth]{fsup/form/mol.pdf}  
    \caption{}
    \end{subfigure}
    \begin{subfigure}{.45\textwidth}
  \centering
    \includegraphics[width=1\linewidth]{fsup/form/eg.pdf}  
     \caption{}
    \end{subfigure}
   \begin{subfigure}{1\textwidth}
    \includegraphics[width=0.32\linewidth]{fsup/form/er1.pdf}  
    \includegraphics[width=0.32\linewidth]{fsup/form/er2.pdf} 
    \includegraphics[width=0.32\linewidth]{fsup/form/er3.pdf} 
      \caption{}
     \end{subfigure}
         \begin{subfigure}{0.5\textwidth}
  \centering
                \includegraphics[width=1\linewidth]{fsup/form/s.pdf} 
                            \caption{} 
    \end{subfigure}
\begin{subfigure}{0.5\textwidth}
       \centering
            \hspace*{0.5cm}  \includegraphics[width=1\linewidth]{fsup/form/w.pdf}  
            \caption{}
\end{subfigure}
\caption{The same plots as in Figure~\ref{fig:m1}, but for the \ce{H2CO} molecule.}
\label{fig:m4}
\end{figure}
\FloatBarrier
\clearpage

\begin{figure}
\begin{subfigure}{0.45\textwidth}
 \centering
\includegraphics[width=0.65\linewidth]{fsup/c2h4/mol.pdf}  
    \caption{}
    \end{subfigure}
    \begin{subfigure}{.45\textwidth}
  \centering
    \includegraphics[width=1\linewidth]{fsup/c2h4/eg.pdf}  
     \caption{}
    \end{subfigure}
   \begin{subfigure}{1\textwidth}
    \includegraphics[width=0.32\linewidth]{fsup/c2h4/er1.pdf}  
    \includegraphics[width=0.32\linewidth]{fsup/c2h4/er2.pdf} 
    \includegraphics[width=0.32\linewidth]{fsup/c2h4/er3.pdf} 
      \caption{}
     \end{subfigure}
         \begin{subfigure}{0.5\textwidth}
  \centering
                \includegraphics[width=1\linewidth]{fsup/c2h4/s.pdf} 
                            \caption{} 
    \end{subfigure}
\begin{subfigure}{0.5\textwidth}
       \centering
            \hspace*{0.5cm}  \includegraphics[width=1\linewidth]{fsup/c2h4/w.pdf}  
            \caption{}
\end{subfigure}
\caption{The same plots as in Figure~\ref{fig:m1}, but for the ethene molecule.}
\label{fig:m5}
\end{figure}
\FloatBarrier
\clearpage

\begin{figure}
\begin{subfigure}{0.45\textwidth}
 \centering
\includegraphics[width=0.9\linewidth]{fsup/met/mol.pdf}  
    \caption{}
    \end{subfigure}
    \begin{subfigure}{.45\textwidth}
  \centering
    \includegraphics[width=1\linewidth]{fsup/met/eg.pdf}  
     \caption{}
    \end{subfigure}
   \begin{subfigure}{1\textwidth}
    \centering
    \includegraphics[width=0.24\linewidth]{fsup/met/er1.pdf}  
    \includegraphics[width=0.24\linewidth]{fsup/met/er2.pdf} 
    \includegraphics[width=0.24\linewidth]{fsup/met/er3.pdf} 
     \includegraphics[width=0.24\linewidth]{fsup/met/er4.pdf} \\
         \centering
     \includegraphics[width=0.24\linewidth]{fsup/met/er5.pdf}  
    \includegraphics[width=0.24\linewidth]{fsup/met/er6.pdf} 
    \includegraphics[width=0.24\linewidth]{fsup/met/er7.pdf}
     \includegraphics[width=0.24\linewidth]{fsup/met/er8.pdf}
     \centering
      \caption{}
     \end{subfigure}
         \begin{subfigure}{0.5\textwidth}
  \centering
                \includegraphics[width=1\linewidth]{fsup/met/s.pdf} 
                            \caption{} 
    \end{subfigure}
\begin{subfigure}{0.5\textwidth}
       \centering
            \hspace*{0.5cm}  \includegraphics[width=1\linewidth]{fsup/met/w.pdf}  
            \caption{}
\end{subfigure}
\caption{The same plots as in Figure~\ref{fig:m1}, but for the methanol molecule.}
\label{fig:m6}
\end{figure}
\FloatBarrier
\clearpage

\begin{figure}
\begin{subfigure}{0.45\textwidth}
 \centering
\includegraphics[width=0.9\linewidth]{fsup/cyclop/mol.pdf}  
    \caption{}
    \end{subfigure}
    \begin{subfigure}{.45\textwidth}
  \centering
    \includegraphics[width=1\linewidth]{fsup/cyclop/eg.pdf}  
     \caption{}
    \end{subfigure}
   \begin{subfigure}{1\textwidth}
    \centering
    \includegraphics[width=0.24\linewidth]{fsup/cyclop/er1.pdf}  
    \includegraphics[width=0.24\linewidth]{fsup/cyclop/er2.pdf} 
    \includegraphics[width=0.24\linewidth]{fsup/cyclop/er3.pdf} 
     \includegraphics[width=0.24\linewidth]{fsup/cyclop/er4.pdf} \\
      \caption{}
     \end{subfigure}
         \begin{subfigure}{0.5\textwidth}
  \centering
                \includegraphics[width=1\linewidth]{fsup/cyclop/s.pdf} 
                            \caption{} 
    \end{subfigure}
\begin{subfigure}{0.5\textwidth}
       \centering
            \hspace*{0.5cm}  \includegraphics[width=1\linewidth]{fsup/cyclop/w.pdf}  
            \caption{}
\end{subfigure}
\caption{The same plots as in Figure~\ref{fig:m1}, but for the cyclopropane molecule.}
\label{fig:m7}
\end{figure}
\FloatBarrier
\clearpage

\begin{figure}
\begin{subfigure}{\textwidth}
 \centering
\includegraphics[width=0.25\linewidth]{fsup/all/h2o.pdf}  
\includegraphics[width=0.35\linewidth]{fsup/all/s_h2o.pdf}  
\includegraphics[width=0.35\linewidth]{fsup/all/w_h2o.pdf}  
   \caption{}
    \end{subfigure}
    \begin{subfigure}{\textwidth}
 \centering
\includegraphics[width=0.25\linewidth]{fsup/all/h2s.pdf}  
\includegraphics[width=0.35\linewidth]{fsup/all/s_h2s.pdf}  
\includegraphics[width=0.35\linewidth]{fsup/all/w_h2s.pdf}  
   \caption{}
    \end{subfigure}
        \begin{subfigure}{\textwidth}
 \centering
\includegraphics[width=0.25\linewidth]{fsup/all/h2co.pdf}  
\includegraphics[width=0.35\linewidth]{fsup/all/s_h2co.pdf}  
\includegraphics[width=0.35\linewidth]{fsup/all/w_h2co.pdf}  
   \caption{}
    \end{subfigure}
            \begin{subfigure}{\textwidth}
 \centering
\includegraphics[width=0.25\linewidth]{fsup/all/c2h4.pdf}  
\includegraphics[width=0.35\linewidth]{fsup/all/s_c2h4.pdf}  
\includegraphics[width=0.35\linewidth]{fsup/all/w_c2h4.pdf}  
   \caption{}
    \end{subfigure}
                \begin{subfigure}{\textwidth}
 \centering
\includegraphics[width=0.25\linewidth]{fsup/all/methanol.pdf}  
\includegraphics[width=0.35\linewidth]{fsup/all/s_methanol.pdf}  
\includegraphics[width=0.35\linewidth]{fsup/all/w_methanol.pdf}  
   \caption{}
    \end{subfigure} 
       \label{fig_all}
    \end{figure}

\begin{figure}\ContinuedFloat
     \clearpage
 \centering
\includegraphics[width=0.25\linewidth]{fsup/all/cyclopropane.pdf}  
\includegraphics[width=0.35\linewidth]{fsup/all/s_cyclopropane.pdf}  
\includegraphics[width=0.35\linewidth]{fsup/all/w_cyclopropane.pdf}  
   \captionof{figure}[Summary of Figs.~\ref{fig:m1}-\ref{fig:m7}]{Summary of Figs.~\ref{fig:m1}-\ref{fig:m7}.}
    \end{figure}
    \FloatBarrier
    
\begin{table}[]
\centering
\footnotesize
\begin{tabular}{llllllllllllll}
                         & HF    & PBE   & BLYP  & TPSS  & B3LYP & MN15  & ωB97  & ωB97XD & LSDA  & PBE {[}D3{]} & MP2   & B2PLYP & PBE0  \\ \hline\hline
\ce{H2O} & 0.450 & 0.129 & 0.173 & 0.094 & 0.014 & 0.038 & 0.009 & 0.013  & 0.138 & 0.129        & 0.000 & 0.003  & 0.004 \\
\ce{H2S}  & 0.139 & 0.103 & 0.098 & 0.021 & 0.006 & 0.066 & 0.016 & 0.002  & 0.099 & 0.104        & 0.019 & 0.002  & 0.002 \\
\ce{H2CO}  & 1.208 & 0.141 & 0.092 & 0.039 & 0.118 & 0.117 & 0.105 & 0.218  & 0.312 & 0.147        & 0.007 & 0.033  & 0.209 \\
ethene                     & 0.436 & 0.118 & 0.064 & 0.037 & 0.079 & 0.085 & 0.091 & 0.123  & 0.251 & 0.120        & 0.029 & 0.052  & 0.100 \\
methanol                   & 0.701 & 0.136 & 0.218 & 0.069 & 0.020 & 0.117 & 0.027 & 0.080  & 0.324 & 0.148        & 0.017 & 0.020  & 0.084 \\
cyclopropane               & 0.408 & 0.137 & 0.118 & 0.032 & 0.059 & 0.172 & 0.118 & 0.116  & 0.522 & 0.143        & 0.057 & 0.061  & 0.159 \\ \hline
mean                       & 0.557 & 0.127 & 0.127 & 0.049 & 0.049 & 0.099 & 0.061 & 0.092  & 0.274 & 0.132        & 0.022 & 0.029  & 0.093 \\
RMS                        & 0.649 & 0.128 & 0.137 & 0.055 & 0.064 & 0.108 & 0.076 & 0.118  & 0.307 & 0.133        & 0.028 & 0.036  & 0.120 
\end{tabular}
\captionof{table}[$\geo$ values for molecules in Figs.~\ref{fig:m1}-~\ref{fig:m7}]{The $\geo$ values for molecules in Figs.~\ref{fig:m1}-~\ref{fig:m7} in kcal/mol. CCSD(T) has been used as a reference in all calculations.  The following basis set have ben used: A'V6Z for the water molecule  aug-cc-pVQZ for \ce{H2S} and ethene, and aug-cc-pVTZ for \ce{H2CO},  methanol, cyclopropane.}
\label{tab_small}
\end{table}
\FloatBarrier
\clearpage

\section{GEO analysis in normal modes (Hessian eigenvectors)}

\begin{figure}
\begin{subfigure}{1\textwidth}
 \centering
\includegraphics[width=0.9\linewidth]{fsup/h2o/m.pdf}  
    \caption{modes that do have contributions to GEO}
    \end{subfigure}
    \begin{subfigure}{0.45\textwidth}
  \centering
    \includegraphics[width=0.4\linewidth]{fsup/h2o/dnm.pdf}  
     \caption{modes that do not have contributions to GEO}
    \end{subfigure}
         \begin{subfigure}{0.45\textwidth}
  \centering
                \includegraphics[width=1\linewidth]{fsup/h2o/b2.pdf} 
                            \caption{Contribution of different normal modes to GEO} 
    \end{subfigure}
\begin{subfigure}{0.5\textwidth}
       \centering
            \hspace*{0.5cm}  \includegraphics[width=1\linewidth]{fsup/h2o/w2.pdf}  
            \caption{Weights of individual normal mode contributions to GEO}
\end{subfigure}
\captionof{table}[Normal-mode GEO analysis for the \ce{H2O molecule}.]{Normal-mode GEO analysis for the \ce{H2O molecule}.   The "1\%" label denotes the GEO errors (or weights) when $ \gamma = 1\%$ (see the main text).}
\label{fig:m12}
\end{figure}
\FloatBarrier
\clearpage

\FloatBarrier
\begin{figure}
\begin{subfigure}{1\textwidth}
 \centering
  \includegraphics[width=0.8\linewidth]{fsup/h2o/wall.pdf}  
\caption{$\geos$ weights for the water molecule in  internal coordinates (left), GEO-active normal modes (right).}
     \end{subfigure}
     \begin{subfigure}{1\textwidth}
      \centering
  \includegraphics[width=0.8\linewidth]{fsup/h2o/vs.pdf}  
  \caption{The change in the bond lengths(left) and bond angle (right) upon scaling GEO active normal modes: $p_{i,\gamma} = (1+\gamma) p_i$ where $p_i$ are {\em accurate} [CCSD(T)] normal modes. The GEO active normal modes are shown in the middle panel.}
      \end{subfigure}
 \caption{Normal-mode vs. internal coordinates GEO analysis for the \ce{H2O molecule}. }
\label{fig:vs}
\end{figure}
The water has 3 normal modes (note again that by normal modes we refer to the eigenvectors of $\mathbf{H}_0$), and two of these are GEO-active. In the left panel of Fig.~\ref{fig:vs}(1) we show the $\geos$ weights in internal coordinates (see Eq.~\ref{eq:gas}), i.e. the $\geo$ contributions from the errors in $\theta$ (bond angle) and $a$ ({\ce OH} bond length). In the right panel of the same figure, we show the $\geo$ weights but given in terms of the errors in normal coordinates (modes), $p_1$ and $p_2$, which correspond to the eigenvectors of the Hessian (see Eq.~\ref{eq:gaNORM}). The third normal mode that leads to an asymmetric {\ce OH} stretch is GEO-inactive (see Fig.~\ref{fig:m12}). 
By the similarity of the patterns in the two panels of Fig.~\ref{fig:vs}(1), we can expect that the changes in $p_1$ mostly affect $a$  and that the changes in  $p_2$ mostly affect $\theta$ (see also the middle panel of Fig.~\ref{fig:vs}(2), where we plot the normal modes.) But, to show explicitly  how the errors in $p_1$ and $p_2$ translate into the errors in $a$ and $\theta$, we consider again the $\gamma$-scaling, and we scale the individual normal coordinates as: $p_i^{\gamma} = \left( 1 + \gamma \right) p_i$. 
From the left panel of Fig.~\ref{fig:vs}(2), we can see that upon scaling of $p_1$ by $1 + \gamma$, $a$ is scaled by the (almost) factor of $\gamma$. At the same time, scaling $p_2$ by $1 + \gamma$ affects very little $a$. 
Note that when all coordinates are scaled by $\gamma$:
\beq \label{eq:gX}
\tilde \G= \left(1+\gamma \right) \G_0~~\text{all angles are unchanged and all bond lengths scaled by } \gamma.
\eeq
Thus, when $p_1$ and $p_2$ are scaled by $1+\gamma$, the angle changes at the same speed in both cases, but in opposite directions, since  the two changes in the angles need to cancel each other out [Eq~\ref{eq:gX}]. This is shown in the right panel of Fig.~\ref{fig:vs}(2). 
Overall, mode 1 contains contributions from both $R$ and $a$, whereas, mode 2 contains mostly contribution from $a$. That is why the weights in the two panels of Fig.~\ref{fig:vs}(1) are similar, but not exactly the same. 
\FloatBarrier

\clearpage

\begin{figure}
\begin{subfigure}{1\textwidth}
 \centering
\includegraphics[width=0.9\linewidth]{fsup/h2s/m.pdf}  
    \caption{modes that do have contributions to GEO}
    \end{subfigure}
    \begin{subfigure}{0.45\textwidth}
  \centering
    \includegraphics[width=0.4\linewidth]{fsup/h2s/dnm.pdf}  
     \caption{modes that do not have contributions to GEO}
    \end{subfigure}
         \begin{subfigure}{0.45\textwidth}
  \centering
                \includegraphics[width=1\linewidth]{fsup/h2s/b2.pdf} 
                            \caption{Contribution of different normal modes to GEO} 
    \end{subfigure}
\begin{subfigure}{0.5\textwidth}
       \centering
            \hspace*{0.5cm}  \includegraphics[width=1\linewidth]{fsup/h2s/w2.pdf}  
            \caption{Weights of individual normal mode contributions to GEO}
\end{subfigure}
\caption{Normal-mode GEO analysis for the \ce{H2S molecule}.}
\label{fig:m22}
\end{figure}
\FloatBarrier
\clearpage

\begin{figure}
\begin{subfigure}{1\textwidth}
 \centering
\includegraphics[width=0.9\linewidth]{fsup/form/m.pdf}  
    \caption{modes that do have contributions to GEO}
    \end{subfigure}
    \begin{subfigure}{0.45\textwidth}
  \centering
    \includegraphics[width=0.4\linewidth]{fsup/form/dnm.pdf}  
     \caption{modes that do not have contributions to GEO}
    \end{subfigure}
         \begin{subfigure}{0.45\textwidth}
  \centering
                \includegraphics[width=1\linewidth]{fsup/form/s2.pdf} 
                            \caption{Contribution of different normal modes to GEO} 
    \end{subfigure}
\begin{subfigure}{0.5\textwidth}
       \centering
            \hspace*{0.5cm}  \includegraphics[width=1\linewidth]{fsup/form/w2.pdf}  
            \caption{Weights of individual normal mode contributions to GEO}
\end{subfigure}
\caption{Normal-mode GEO analysis for the \ce{H2CO molecule}.}
\label{fig:m42}
\end{figure}
\FloatBarrier
\clearpage

\begin{figure}
\begin{subfigure}{1\textwidth}
 \centering
\includegraphics[width=0.9\linewidth]{fsup/c2h4/m.pdf}  
    \caption{modes that do have contributions to GEO}
    \end{subfigure}
    \begin{subfigure}{0.45\textwidth}
  \centering
    \includegraphics[width=0.4\linewidth]{fsup/c2h4/dnm.pdf}  
     \caption{modes that do not have contributions to GEO}
    \end{subfigure}
         \begin{subfigure}{0.45\textwidth}
  \centering
                \includegraphics[width=1\linewidth]{fsup/c2h4/b2.pdf} 
                            \caption{Contribution of different normal modes to GEO} 
    \end{subfigure}
\begin{subfigure}{0.5\textwidth}
       \centering
            \hspace*{0.5cm}  \includegraphics[width=1\linewidth]{fsup/c2h4/w2.pdf}  
            \caption{Weights of individual normal mode contributions to GEO}
\end{subfigure}
\caption{Normal-mode GEO analysis for the \ce{ethene} molecule.}
\label{fig:m52}
\end{figure}
\FloatBarrier
\clearpage

\section{Details on the $\gamma$ expansion and $\K$ values for molecular sets} 

The $\K$ molecule-specific constant (absolute GEO scale) is given by:
\ben \label{eq:k1}
\K= \G_0^\intercal  \H_0  \G_0,
\een
and thereby can be easily calculated once the force constants (in e.g. Cartesian coordinates) are available. 
In analogy to approximating $\geo$ by $\geop$, we can calculate $\K$ more cheaply from approximate minima and force constants:
\ben \label{eq:k2}
\K^{\prime}=  \tilde{\G}^\intercal  \tilde{\H}  \tilde{\G}.
\een
Furthermore, in analogy to $\geos$ approximation to $\geo$, we can also define $\K^{simple}$ as:
\ben  \label{eq:ks}
\K \approx \K^{simple}=\sum_{i}^{\rm all~bonds} K^{simple,i}  = \sum_{i}^{\rm all~bonds}  R_i^2~f_{ii} ,
\een
where $R_i$-s are the bond lengths and $f_{ii}$ are the force constants for each of the bonds. 
Note that there is no angle contribution to the r.h.s. of Eq~\ref{eq:ks}, since by the $\gamma$-scaling (see the main text),  angles remain unchanged. 

In Table~S6, we show the accurate $\K$ and approximate $\K'$ values. From this table, we can see that even HF $\K'$ are decently accurate and already B3LYP ones are in an excellent agreement with $\K$. 
For the same set of molecules, $E^{simple}$ and contributions from individual bonds, we show in Fig.~\ref{fig:simplek}.
For other organic molecules in the paper, we give the $\K$ values in Tables~\ref{tab:k1} and ~S8}. 
Finally, in Table~\ref{tab:kn}, we show the $\K$ values for rare gas dimers, and we can see that these $\K$ values are by an order of magnitude smaller than those for covalently bonded molecules.

\begin{table}[!h]
	\centering
\begin{tabular}{ccccccccc}
 \text{} & \text{CCSD(T)} & \text{MP2} & \text{B2PLYP} & \text{B3LYP} & \text{PBE0} &
\text{TPSS} & \text{PBE} & \text{HF} \\  \hline
\text{h2o} & 0.221 & 0.222 & 0.220 & 0.219 & 0.224 & 0.211 & 0.211 & 0.247 \\
\text{h2s} & 0.220 & 0.228 & 0.221 & 0.215 & 0.221 & 0.213 & 0.209 & 0.239 \\
\text{methanol} & 0.566 & 0.572 & 0.567 & 0.563 & 0.579 & 0.547 & 0.545 & 0.629 \\
\text{cyclopropane} & 0.989 & 1.007 & 0.998 & 0.99 & 1.012 & 0.972 & 0.966 & 1.076 \\
\text{ch3cl} & 0.452 & 0.465 & 0.452 & 0.443 & 0.459 & 0.436 & 0.435 & 0.488 \\
\text{ch3f} & 0.463 & 0.469 & 0.463 & 0.458 & 0.47 & 0.447 & 0.443 & 0.511 \\
\text{c2h3cl} & 0.727 & 0.742 & 0.736 & 0.733 & 0.754 & 0.717 & 0.714 & 0.81 \\
\text{c2h4} & 0.634 & 0.643 & 0.643 & 0.642 & 0.651 & 0.629 & 0.623 & 0.699 \\
\text{c-c2h2f2} & 0.848 & 0.858 & 0.856 & 0.856 & 0.881 & 0.828 & 0.823 & 0.971 \\
\text{c2h3f} & 0.735 & 0.745 & 0.744 & 0.744 & 0.761 & 0.724 & 0.719 & 0.828 \\
\text{imine} & 0.555 & 0.557 & 0.564 & 0.569 & 0.58 & 0.551 & 0.547 & 0.636 \\
\text{h2co} & 0.485 & 0.485 & 0.489 & 0.497 & 0.508 & 0.48 & 0.476 & 0.563 \\
\text{formic} & 0.725 & 0.724 & 0.723 & 0.731 & 0.753 & 0.698 & 0.697 & 0.842 \\
\text{hcocl} & 0.594 & 0.595 & 0.592 & 0.599 & 0.623 & 0.573 & 0.573 & 0.696 \\ \hline \hline
\text{MAE} & / & 0.007 & 0.005 & 0.007 & 0.019 & 0.013 & 0.017 & 0.073 \\
\text{RMS} & / & 0.009 & 0.006 & 0.008 & 0.021 & 0.015 & 0.018 & 0.079 \\
\text{MARE ($\%$)} &/ & 1.3 & 0.7 & 1.3 & 3. & 2.5 & 3.1 & 12.2 \\
	\end{tabular}
	\captionof{table}[$\K$ values for molecules considered in Fig. 1b.] {Accurate $\K$ values [CCSD(T)]   in $10^4$ kcal/mol, and approximate $\K'$ values computed from Eqs.~\ref{eq:k1} and~\ref{eq:k2}, respectively, for the set of molecules considered in Figs. 1b.}
			\label{tab:ktabx}
\end{table}

\begin{figure}
	\centering
	\includegraphics[width=4cm]{fsup/K/1.pdf}
	\includegraphics[width=4cm]{fsup/K/2.pdf}
	\includegraphics[width=4cm]{fsup/K/3.pdf}
	\includegraphics[width=4cm]{fsup/K/4.pdf}\\
	\centering
	\includegraphics[width=4cm]{fsup/K/5.pdf}
	\includegraphics[width=4cm]{fsup/K/6.pdf}
	\includegraphics[width=4cm]{fsup/K/7.pdf}
	\includegraphics[width=4cm]{fsup/K/8.pdf}\\
	\centering
	\includegraphics[width=4cm]{fsup/K/9.pdf}
	\includegraphics[width=4cm]{fsup/K/10.pdf}
	\includegraphics[width=4cm]{fsup/K/11.pdf}
	\includegraphics[width=4cm]{fsup/K/12.pdf}\\
	\centering
	\includegraphics[width=4cm]{fsup/K/13.pdf}
	\includegraphics[width=4cm]{fsup/K/14.pdf}
	\captionof{figure}[$\K^{simple}$ bond-decomposition for the set of molecules considered in  Fig. 1b.]{$\K$ values in $10^4$ kcal/mol  (a value in black in the top-right corner of each panel), $\K^{simple}$ values (a value in blue the top-right corner of each panel), and $\K^{simple}$ bond-decomposition of Eq.~\ref{eq:ks}
		for the set of molecules considered in  Fig. 1b. } 
		\label{fig:simplek}
\end{figure}

\begin{table}[!h]
	\centering
	\begin{tabular}{cc}
		\text{acetaldehyde} & 0.842 \\
		\text{acetic} & 1.086 \\
		\text{c2h3f} & 0.735 \\
		\text{c2h5f} & 0.819 \\
		\text{ch2ch} & 0.749 \\
		\text{ch2nh2} & 0.598 \\
		\text{ch3nh} & 0.543 \\
		\text{ch3nh2} & 0.660 \\
		\text{ethanol} & 0.917 \\
		\text{formic} & 0.73 \\
		\text{methanol} & 0.568 \\
		\text{propane} & 1.066 \\
		\text{propene} & 0.982 \\
		\text{propyne} & 0.877 \\
	\end{tabular}
	\captionof{table}[$\K$ values for molecules considered in Fig. 1a(top)]{$\K$ values in $10^4$ kcal/mol for individual molecules for molecules considered in Fig. 1a(top) \computed at the CCSD(T) level of theory. 
		For sufficiently small $\gamma$, $\K$ can be computed by finite difference from: 
		$\K=\frac{2 \geo ^ \gamma}{\gamma^2}$, and we have used here $\gamma=0.5 \%$. We have found that this $\gamma$ value is small enough to stay within the harmonic approximation, but still not too small to lose numerical accuracy.}
	\label{tab:k1}
\end{table}

\begin{table}[!h]
	\centering
\begin{tabular}{cc}
	mol. $\#$ & K \\
		1 & 2.544 \\
	2 & 1.695 \\
	3 & 2.229 \\
	4 & 2.386 \\
	5 & 2.083 \\
	6 & 2.483 \\
	7 & 2.505 \\
	8 & 1.693 \\
	9 & 2.387 \\
	10 & 2.009 \\
	11 & 2.457 \\
	12 & 2.339 \\
	13 & 2.738 \\
	14 & 3.208 \\
	15 & 3.715 \\
	16 & 3.404 \\
\end{tabular}
	\label{tab:k2}
		\captionof{table}[$\K$ values for individual molecules considered in Fig. 1a (bottom)]{$\K$ values in $10^4$ kcal/mol obtained as described in the caption Table~\ref{tab:k1}, but with B2PLYP/aug-cc-pVTZ as a reference. The order of molecules is shown in Fig.~\ref{fig_grid}.}
		\end{table}

\begin{table}[!h]
		\centering
\begin{tabular}{cc}
		\ce{He2} & 0.000154 \\
		\ce{Ne2} & 0.000724 \\
		\ce{Ar2} & 0.002164 \\
		\ce{HeNe} & 0.000301 \\
		\ce{NeAr} & 0.001079 \\
\end{tabular}
\captionof{table}[$\K$ values  in $10^4$ kcal/mol for noble gas dimers]{$\K$ values in $10^4$ kcal/mol computed from the CCSD(T) force constants for the noble gas dimers (Eq.~\ref{eq:k1}). For the basis set informations, see Fig.~\ref{fig:rg}.}
\label{tab:kn}
\end{table}

\clearpage

\section{Statistical measures for the quality of approximate geometries of the methylamine molecule} 
\begin{figure}[htb]
\includegraphics[width=0.98\columnwidth]{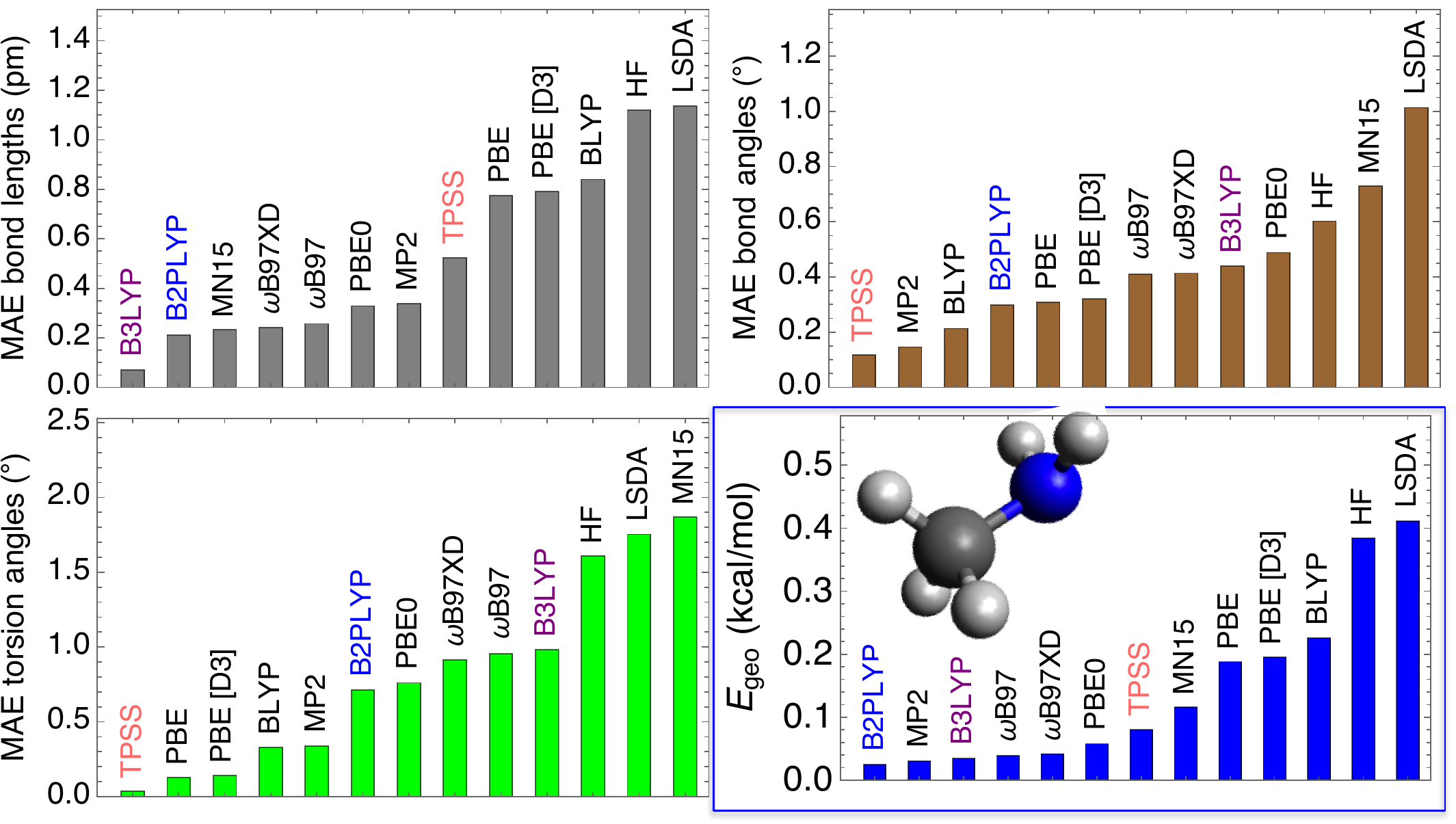}
\captionof{figure}[$\geo$ and other statistical measures for the quality of approximate methylamine geometries] {Mean absolute errors (MAE) in (i) calculated bond lengths; 
(ii) bond angles; (iii) torsion angles (iv) GEO errors obtained from various approximate methods 
for the methylamine molecule.}
\label{fig:amine2}
\end{figure}

In addition to the measures in Figure~\ref{fig:amine2}, in Figure~\ref{fig_amine2}, we show other approximate measures for the geometry of the methylamine molecule.  For comparison, we also show in Figure~\ref{fig_aa} the errors in atomization energies obtained with the same methods. 

\begin{figure*}
\centering
\includegraphics[width=5cm]{fsup/f/a_MSE_b.pdf}
\includegraphics[width=5cm]{fsup/f/a_MSE_a.pdf}
\includegraphics[width=5cm]{fsup/f/a_MSE_t.pdf}\\
\centering
\includegraphics[width=5cm]{fsup/f/a_MAE_C.pdf}
\includegraphics[width=5cm]{fsup/f/a_MAE_N.pdf}
\includegraphics[width=5cm]{fsup/f/a_MAE_CN.pdf}
\captionof{figure}[Other statistical measures for the quality of approximate geometries of the methylamine molecule]{Other statistical measures for the quality of the approximate geometries of the methylamine molecule (see Figure 3 of the main text):
(i) means signed error (MSE) for the bond lengths; (ii): MSE for the bond angles; (iii) MSE for the torsion angles
(iv) mean absolute error (MAE) for the distances between the C atom and the other atoms in the molecule
(v)  MAE for the distances between the N atom and the other atoms in the molecule 
(vi)  MAE for the distances between the CN midbond point and the atoms in the molecule.}
\label{fig_amine2}
\end{figure*}

\FloatBarrier

\begin{figure}
\centering
\includegraphics[width=9cm]{fsup/f/a_atom.pdf}
\captionof{figure}[Atomization energy errros for the methylamine molecule] {Absolute errors for the atomization energies of the methylamine molecule obtained from different methods. The LSDA and HF errors are divided by the factor of $4.5$}
\label{fig_aa}
\end{figure}

\FloatBarrier

\section{GEO analysis for noble gas dimers}
\FloatBarrier
\begin{figure}
                       \begin{subfigure}{1\textwidth}
                          \centering
    \includegraphics[width=0.32\linewidth]{fsup/ncv/ne21.pdf}  
    \includegraphics[width=0.32\linewidth]{fsup/ncv/ne22.pdf} 
       \includegraphics[width=0.32\linewidth]{fsup/ncv/ne23.pdf} 
      \caption{{\ce Ne2} [0.09 kcal/mol]}
     \end{subfigure}
     \centering
                            \begin{subfigure}{1\textwidth}
                                \includegraphics[width=8cm]{fsup/ncv/ne2curs.pdf}  
                                \end{subfigure}
                                \captionof{figure}[Geo analysis for the neon dimer.]{Top panel: Geo analysis for the neon dimer. For the basis set information, see Figure~\ref{fig:rg}. Accurate binding energy of the dimer is shown in the square brackets in the subcaption. The aug-cc-pV5Z basis set has been used in all calculations.
                                In the left top panel, the $\Delta R$ position of approximations are not accurate when $\Delta R > 1\angstrom$.  In these cases, the approximations typically give unphysical binding curves with no minimum and only infimum at $R \to \infty$. Bottom panel: Various dissociation curves for the neon dimer. }
\label{fig:ne2geo}
\end{figure}

\begin{figure}
\includegraphics[width=8cm]{fsup/ncv/ne2_cp.pdf} 
\label{fig:ne2bsse}
\captionof{figure}[Counterpoise corrected and counterpoise uncorrected binding curves for the neon dimer]{Counterpoise (CP) corrected (dashed lines) and counterpoise uncorrected binding curves for the neon dimer (solid lines) obtained from various approximations. Detailed analysis of Baerends and co-workers have shown that for small dimers bonded by dispersion, CP-uncorrected curves with a finite basis set are in a better agreement with the  "exact" (i.e. no errors due to a finite basis set) curve  curve than its CP-corrected counterpart (see Refs.~\cite{cbp1,cbp2}). Thus, in the main text and in Figure~\ref{fig:ne2geo}, we use CP-uncorrected curves. Nonetheless, due to a large basis set used here [aug-cc-pV5Z], we can see from this figure that the energy difference between curves with and without the CP is small, even in the CCSD(T) case and nearly negligible in the case of DFT curves.}
\end{figure}
\FloatBarrier

\begin{figure}
   \begin{subfigure}{1\textwidth}
    \includegraphics[width=0.32\linewidth]{fsup/ncv/he21.pdf}  
    \includegraphics[width=0.32\linewidth]{fsup/ncv/he22.pdf} 
       \includegraphics[width=0.32\linewidth]{fsup/ncv/he23.pdf} 
      \caption{{\ce He2} [0.02 kcal/mol]}
     \end{subfigure}
        \begin{subfigure}{1\textwidth}
    \includegraphics[width=0.32\linewidth]{fsup/ncv/hene1.pdf}  
    \includegraphics[width=0.32\linewidth]{fsup/ncv/hene2.pdf} 
       \includegraphics[width=0.32\linewidth]{fsup/ncv/hene3.pdf} 
      \caption{{\ce HeNe} [0.05 kcal/mol]}
     \end{subfigure}
             \begin{subfigure}{1\textwidth}
    \includegraphics[width=0.32\linewidth]{fsup/ncv/near1.pdf}  
    \includegraphics[width=0.32\linewidth]{fsup/ncv/near2.pdf} 
       \includegraphics[width=0.32\linewidth]{fsup/ncv/near3.pdf} 
      \caption{{\ce NeAr} [0.09 kcal/mol]}
     \end{subfigure}
                  \begin{subfigure}{1\textwidth}
    \includegraphics[width=0.32\linewidth]{fsup/ncv/ar21.pdf}  
    \includegraphics[width=0.32\linewidth]{fsup/ncv/ar22.pdf} 
       \includegraphics[width=0.32\linewidth]{fsup/ncv/ar23.pdf} 
      \caption{{\ce Ar2} [0.27 kcal/mol]}
     \end{subfigure}
         \captionof{figure}[The GEO analysis for noble gas dimers]{The GEO analysis for noble gas dimers. The aug-cc-pV5Z basis has been used for {\ce Ne2} and {\ce He2}, and aug-cc-pVQZ has been used for all other dimers. CCSD(T) has been used as a reference for all dimers. In the left panels, the $\Delta R$ position of approximations are not accurate when $\Delta R > 1 \angstrom$. In these cases, the approximations typically give unphysical binding curves with no minimum and only infimum at $R \to \infty$. }
\label{fig:rg}
\end{figure}
\FloatBarrier
\clearpage

\section{Details on the results of Figure~\ref{fig:s66} [S66 results]}
The S66x8 set contains 66 dimers bonded by noncovalent interactions.\cite{RezRilHob-JCTC-11}
For each complex and each approximate method we do single point calculations to obtain 8 datapoints for different separations between fragments along the dissociation curve of a given complex. 
The geometries at different separations between fragments in a complex have been taken from the S66x8 dataset.\cite{RezRilHob-JCTC-11} The CCSD(T)/CBS (used as a reference here), MP2/CBS, CCSD/CBS and SCS-CCSD/CBS binding energies have been taken from the original S66x8 dataset.\cite{RezRilHob-JCTC-11,rezac}  All other binding energy we obtain from the G16, by performing counterpoise corrected calculations within the aug-cc-pVTZ basis set.
From these 8 datapoints (for each complex and for each approximate method) we use a spline interpolation to construct a dissociation curve. 
The $\geo$ error for a given approximate method  
is then calculated by the difference between the CCSD(T) binding energies at the approximate minimum and at the CCSD(T) minimum (see Eq.~\ref{eq:err1}). 
In principle, in addition to finding a minimum intermolecular separation between fragments, 
we should have also optimized the fragments with a given approximate method. However, we assume here that this has a minor effect on the 
$\geo$ contribution to the binding energies errors of the S66 complexes. 

The same plots shown in Figure~\ref{fig:s66} in the main text are shown here with other approximate methods. Namely, in Figure~\ref{fig_D3} we show the $\left| \geo  \right|$ vs. $\left| \Delta E \right|$ plots (as in Figure~\ref{fig:s66} of the main text) for selected DFT functionals
with the D3 empirical correction. The same plots are shown in Figure~\ref{fig_cc} for CCSD/CBS and SCS-CCSD/CBS. 
The same plots obtained from the PBE functional with and without the D3 correction is shown in Figure~\ref{fig_pbe}. 
\FloatBarrier
\begin{figure*}
\centering
\includegraphics[width=8.5cm]{fsup/UB3LYP_GD3.pdf}
\includegraphics[width=8.5cm]{fsup/UM06_GD3.pdf}
\centering
\includegraphics[width=8.5cm]{fsup/UBLYP_GD3.pdf}
\includegraphics[width=8.5cm]{fsup/UwB97XD.pdf}
\caption{$\geo$ (y-axis) vs. $\left | \Delta E \right | $ (x-axis)  errors in kcal/mol for different methods and for the S66 dataset.}
\label{fig_D3}
\end{figure*}

\FloatBarrier

\begin{figure}
\centering
\includegraphics[width=8.5cm]{fsup/ccsd.pdf}
\includegraphics[width=8.5cm]{fsup/scs_ccsd.pdf}
\caption{$\geo$ (y-axis) vs. $\left | \Delta E \right | $ (x-axis)  errors in kcal/mol for different methods and for the S66 dataset.}
\label{fig_cc}
\end{figure}

\FloatBarrier

\begin{figure}
\centering
\includegraphics[width=8.5cm]{fsup/UPBEPBE.pdf}
\includegraphics[width=8.5cm]{fsup/UPBEPBE_GD3.pdf}
\caption{$\geo$ (y-axis) vs. $\left | \Delta E \right | $ (x-axis)  errors in kcal/mol for different methods and for the S66 dataset.}
\label{fig_pbe}
\end{figure}

\begin{table*}[!h]
\begin{array}{cccccccccc}
 \text{   } & \text{MP2/CBS} & \text{B2PLYP} & \text{B3LYP[D3]} & \text{PBE0[D3]} & \text{PBE[D3]} & \text{MN15} & \text{wB97XD} &
   \text{PBE} & \text{CCSD(T)} \\
 \text{MP2/CBS} & \textcolor{blue}{0.51} & 0.26 & 0.45 & 0.43 & 0.38 & 0.47 & 0.46 & 0.28 & \textcolor{red}{0.47} \\
 \text{B2PLYP} & 1.63 & \textcolor{blue}{1.34} & 1.45 & 1.43 & 1.38 & 1.52 & 1.51 & 1.40 &  \textcolor{red}{1.46} \\
 \text{B3LYP[D3]} & 0.44 & 0.34 & \textcolor{blue}{0.43} & 0.42 & 0.41 & 0.41 & 0.42 & 0.53 &  \textcolor{red}{0.43} \\
 \text{PBE0[D3]} & 0.51 & 0.43 & 0.48 & \textcolor{blue}{0.49} & 0.48 & 0.45 & 0.47 & 0.54 &  \textcolor{red}{0.48} \\
 \text{PBE[D3]} & 0.48 & 0.4 & 0.44 & 0.45 & \textcolor{blue}{0.46} & 0.40 & 0.41 & 0.44 &  \textcolor{red}{0.44} \\
 \text{MN15} & 0.56 & 0.34 & 0.59 & 0.58 & 0.53 & \textcolor{blue}{0.6} & 0.60 & 0.34 &  \textcolor{red}{0.59} \\
 \text{wB97XD} & 0.42 & 0.25 & 0.47 & 0.45 & 0.41 & 0.47 & \textcolor{blue}{0.48} & 0.38 &  \textcolor{red}{0.46} \\
 \text{PBE} & 2.29 & 1.77 & 2.04 & 1.98 & 1.90 & 2.14 & 2.12 & \textcolor{blue}{1.71} & \textcolor{red}{2.06} \\
\end{array}
\captionof{table}[]{MAE of different methods (rows) at different minima of S66x8 binding curves (columns) for the S66 dataset. The error is defined as: $E_X [R_Y] - E_0$, where $X$ is a method in rows and Y is the method in columns, and $E_0$ is the binding energy at the CCSD(T) minimum. The S66x8 binding curves from CCSD(T)/CBS (used as a reference here) and MP2/CBS have been taken from the original S66x8 dataset.\cite{RezRilHob-JCTC-11,rezac}  All other binding curves have been obtained from counterpoise corrected calculations within the aug-cc-pVTZ basis set. The minimum of each binding curve has been found numerically after the interpolation of 8 datapoints for each of the S66 complex (see details above). }
\label{tab_s66_2}
\end{table*}

\begin{figure}
	\centering
	\includegraphics[width=8.5cm]{fsup/28.pdf}
	\caption{Benzene-uracil binding curves (interpolated from the 8 datapoints along the dissociation curve).}
	\label{fig_28}
\end{figure}

\clearpage


\bibliographystyle{unsrt}
\bibliography{all}

\label{page:end}